\documentclass[%
reprint,
 amsmath,amssymb,
 aps,superscriptaddress
]{revtex4-1}
\usepackage{graphicx}% Include figure files
\usepackage{dcolumn}% Align table columns on decimal point
\usepackage{bm}% bold math
\usepackage{braket}
\usepackage{color}
\usepackage{hyperref}
\begin{document}

\title{Appearance of hinge states in second-order topological insulators via the cutting procedure
}

\author{Yutaro Tanaka}
\affiliation{
 Department of Physics, Tokyo Institute of Technology, 2-12-1 Ookayama, Meguro-ku, Tokyo 152-8551, Japan
}
\author{Ryo Takahashi}
\affiliation{
 Department of Physics, Tokyo Institute of Technology, 2-12-1 Ookayama, Meguro-ku, Tokyo 152-8551, Japan
}
\author{Shuichi Murakami}
\affiliation{
 Department of Physics, Tokyo Institute of Technology, 2-12-1 Ookayama, Meguro-ku, Tokyo 152-8551, Japan
}
\affiliation{
 TIES, Tokyo Institute of Technology, 2-12-1 Ookayama, Meguro-ku, Tokyo 152-8551, Japan
}

\begin{abstract}
In recent years, second-order topological insulators have been proposed as a new class of topological insulators.
Second-order topological insulators are materials with gapped bulk and surfaces, but with topologically protected gapless states at the intersection of two surfaces. These gapless states are called hinge states. 
In this paper, we give a general proof that any insulators with inversion symmetry and gapped surface in class A always have hinge states when the $\mathbb{Z}_{4}$ topological index $\mu_{1}$ is $\mu_{1}=2$.
We consider a three-dimensional insulator whose boundary conditions along two directions change by changing the hopping amplitudes across the boundaries. 
We study behaviors of gapless states through continuously changing boundary conditions along the two directions, 
and reveal that the behaviors of gapless states result from the $\mathbb{Z}_{4}$ strong topological index. 
From this discussion, we show that gapless states inevitably appear at the hinge of a three-dimensional insulator with gapped surfaces when the strong topological index is
$\mathbb{Z}_{4}=2$ and the weak topological indices are $\nu_{1}=\nu_{2}=\nu_{3}=0$.
\end{abstract}

\maketitle

%\tableofcontents

\section{INTRODUCTION}
A topological insulator (TI) is a material with the insulating bulk, but with topologically protected gapless surface or edge states
\cite{PhysRevLett.95.146802,PhysRevLett.95.226801,bernevig2006quantum,PhysRevLett.98.106803,PhysRevB.76.045302,PhysRevB.78.045426,PhysRevB.78.195424,RevModPhys.82.3045,RevModPhys.83.1057}.
TIs are classified in terms of $\mathbb{Z}_{2}$ topological invariants and their gapless surface states are 
topologically protected.
Two- and three-dimensional TIs have one-dimensional (1D) and two-dimensional (2D) topological gapless states,  respectively.
Namely, topological nature in the $n$-dimensional bulk of the system is associated with ($n-1$)-dimensional gapless states in TIs.

In recent years, second-order topological insulators (SOTI) have been proposed as a new class of topological insulators
 \cite{PhysRevLett.108.126807,PhysRevLett.110.046404,PhysRevB.92.085126, benalcazar2017quantized, PhysRevB.96.245115,PhysRevLett.119.246402, fang2017rotation,PhysRevB.97.241405, schindler2018higher,  PhysRevLett.119.246401,PhysRevLett.120.026801,PhysRevLett.121.116801,PhysRevB.97.155305, PhysRevB.97.241402,  PhysRevB.97.205136, schindler2018higherbismuth, PhysRevB.98.205129, wang2018higher, PhysRevLett.122.256402,ezawa2019scientificrepo,PhysRevB.98.201114, PhysRevB.97.205135, PhysRevB.98.081110, PhysRevB.98.245102,PhysRevX.9.011012,okugawa2019second,PhysRevB.99.041301,yue2019symmetry,zhang2019second,PhysRevLett.122.076801,PhysRevB.98.205147,PhysRevLett.121.196801,PhysRevB.98.235102,PhysRevLett.123.016805,PhysRevLett.123.073601,serra2018observationnature555,peterson2018quantizedNature7695,imhof2018topolectricalnatphys,PhysRevB.99.195431,PhysRevLett.123.016806,sheng2019two,agarwala2019higher,PhysRevLett.123.036802,chen2019higher,PhysRevB.98.035147,wieder2018axioninsulatorpump}. In three dimensions, SOTIs are insulating both in the bulk and in the surface. However, 
they have anomalous gapless states at an intersection of two surfaces, called hinge states.
In the SOTIs, the topological nature of $n$-dimensional bulk manifests itself not as $(n-1)$- but as $(n-2)$-dimensional gapless states.
Among various classes of SOTIs, one class of SOTIs is protected by inversion symmetry
 \cite{wang2018higher,ezawa2019scientificrepo,PhysRevLett.122.256402, PhysRevB.98.205129,PhysRevB.97.205136,schindler2018higherbismuth}, and this class of SOTIs is characterized by a $\mathbb{Z}_{4}$ index of symmetry-based indicators \cite{schindler2018higherbismuth, wang2018higher,ezawa2019scientificrepo, PhysRevX.7.041069, po2017symmetry,PhysRevX.8.031070, PhysRevB.98.115150, PhysRevLett.122.256402, PhysRevB.98.205129,tang2019comprehensive,tang2019efficient}. 
Appearance of the hinge states in a SOTI is usually understood in term of the surface Dirac Hamiltonian with a symmetry-respecting mass term. The surface energy spectrum is gapped by adding this mass term. However, due to symmetry constraint, the mass term changes its sign depending on the surface direction. Therefore, at the intersection of two surfaces having mass terms with opposite signs, the mass terms can be regarded as zero. This allows behaviors of electrons at the hinge to be represented as massless Dirac hamiltonian and the energy spectrum becomes gapless at the hinge.

As described above, appearance of a hinge state is topologically protected by symmetry. However, in the above discussion with the surface Dirac Hamiltonian, we cannot directly explain how the $\mathbb{Z}_{4}$ index of symmetry-based indicators is related to the hinge state. 
In this paper, our purpose is to show the emergence of the gapless hinge states when the $\mathbb{Z}_{4}$  index of symmetry-based indicators is nontrivial without relying upon specific models. 
The previous work discussing the connection between symmetry-based indicators and hinge states \cite{PhysRevX.8.031070}, is based on the $\boldsymbol{k}\cdot \boldsymbol{p}$ Dirac Hamiltonian. Therefore, this approach cannot be applied to systems whose surfaces are not described by Dirac model. In order to complete the proof, it is necessary to establish a theory on the connection between the $\mathbb{Z}_{4}$ index and hinge states for general systems.
There have been studies based on tight-binding models of SOTIs with a nontrivial $\mathbb{Z}_{4}$ index of symmetry-based indicators \cite{schindler2018higherbismuth, PhysRevB.98.205129, wang2018higher, PhysRevLett.122.256402,ezawa2019scientificrepo}.
However, this argument does not lead to a general proof that hinge states appear generally in any model with  a nontrivial $\mathbb{Z}_{4}$ index of symmetry-based indicators. 

In this paper, we propose a new method to understand the hinge state only from the $\mathbb{Z}_{4}$ index of symmetry-based indicators.
This method is applicable to a broad range of systems.
In this method, we change boundary conditions in two directions by changing hopping amplitude across the boundaries. When the hopping amplitudes across the two boundaries become zero, the system is cut along two planes, giving rise to a hinge. Then by tracing the spectral flow along the change, we can see whether and how hinge states appear.
From this discussion, we show that when the 
$\mathbb{Z}_{4}$ topological index $\mu_{1}$ is $\mu_{1}=2$ for class A, gapless states appear inevitably at the hinges of three-dimensional insulators with inversion symmetry.
We note that the gapless states may not be localized at the hinges if surfaces are gapless. Therefore we restrict ourselves to the case with no gapless surface states throughout the present paper. In the main text of this paper, we consider systems in class A, and we extend our theory to systems in class AII in Appendix \ref{section:timereversal}.

A similar method with changing the hopping amplitude across only one boundary has been applied to characterize TIs \cite{PhysRevB.78.045426} and SOTIs \cite{bulkhinge}, and this method is called cutting procedure. 
In Ref.~\cite{bulkhinge}, the boundary condition is changed only along one direction, in contrast with the present paper. Through this change the three-dimensional system is related with a two-dimensional slab. Then we show that the indicators, which characterize three-dimensional inversion-symmetric SOTIs, are directly related to the indicators of the two-dimensional inversion-symmetric systems, i.e. the Chern number parity. In the present paper, we study the spectral flows in the band gap, i.e. the behaviors of gapless states through continuously changing the boundary conditions along the two directions.  In addition, we find the spectral flows related to appearance of the hinge states.

This paper is organized as follows. In Sec.~\ref{topological index and cutting procedure}, we explain the $\mathbb{Z}_{4}$ topological index and cutting procedure. In addition, we show appearance of hinge states by the applying the cutting procedure to one of the models with $\mathbb{Z}_{4}=2$. In Sec.~\ref{section:Tight-binding model}, we confirm our theory in Sec.~\ref{topological index and cutting procedure} by calculations on a tight-binding model of a SOTI. In Sec.~\ref{sec:positionofhingestates}, we discuss which of the hinges have hinge states. Our conclusion is given in Sec.~\ref{section:conclusion}.

%%section2
\section{\label{topological index and cutting procedure}$\mathbb{Z}_{4}$ topological index and cutting procedure}
In this section, we will establish the relationship between the hinge state and $\mathbb{Z}_{4}$ topological index by cutting procedure. 
\subsection{Strong $\mathbb{Z}_{4}$ index and weak $\mathbb{Z}_{2}$ indices}

\begin{figure}
\includegraphics{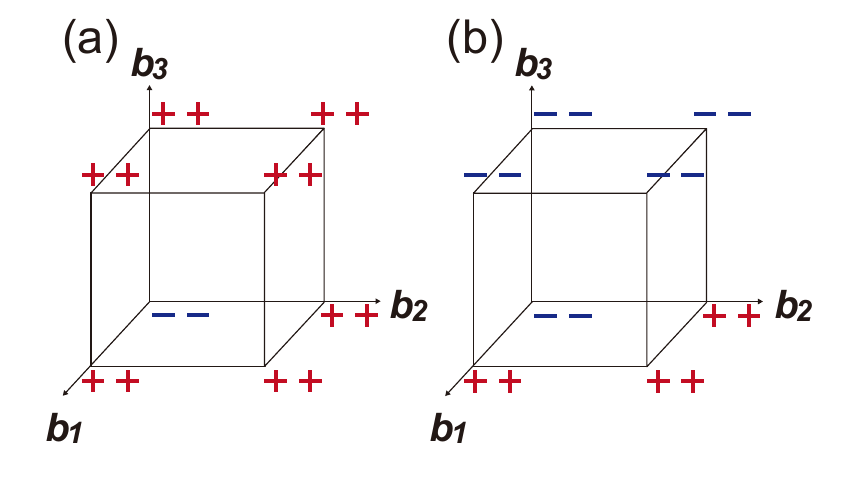}
\caption{\label{zfourparitypicture}(Color online) Parity eigenvalues at TRIM.~(a, b) Two examples of parity eigenvalues at TRIM to realize $(\nu_{1}, \nu_{2}, \nu_{3}, \mu_{1})=(0,0,0,2)$. They are transformed to each other by a shift of an inversion center by $\boldsymbol{a}_{3}/2$.}
\end{figure}
We consider a noninteracting centrosymmetric system on a three-dimensional lattice in class A, one of the Altland-Zirnbauer symmetry classes [\onlinecite{PhysRevB.55.1142}].
For three-dimensional systems, there are eight time-reversal invariant momenta (TRIM) denoted by $\Gamma_{j}$.
The eight TRIM $\Gamma_{j}$ can be indexed by three integers $n_{l}=0,1$ defined mod 2,
\begin{equation}
\Gamma_{j=(n_{1},n_{2},n_{3})}=\frac{1}{2}(n_{1}\boldsymbol{b}_{1}+n_{2}\boldsymbol{b}_{2}+n_{3}\boldsymbol{b}_{3}),
\end{equation}
where $\boldsymbol{b}_{l}$ are primitive reciprocal lattice vectors.
According to Ref. [\onlinecite{PhysRevB.98.115150}], the symmetry indicator for class A is found to be $X_{\rm BS}=\mathbb{Z}_{2}\times \mathbb{Z}_{2}\times \mathbb{Z}_{2}\times \mathbb{Z}_{4}$. 
The three factors of $\mathbb{Z}_{2}$ are the weak topological indices
\begin{equation}\label{weakz2index}
\nu_{a}\equiv \sum_{\Gamma_{j}:{\rm TRIM}\land n_{a}=1}n_{-}(\Gamma_{j})\ \ \ ({\rm mod}\ 2)\ (a=1,2,3),
\end{equation}
where $n_{-}(\Gamma_{i})$ is the number of occupied states with odd parity at the TRIM $\Gamma_{j}$, and the summation is taken over the TRIM on the plane $n_{a}=1$. 
The factor of $\mathbb{Z}_{4}$ is the strong topological index, defined as 
\begin{align}\label{z4index}
\mu_{1}\equiv 
&\frac{1}{2}\sum_{\Gamma_{j}:{\rm TRIM}}\Bigl(n_{+}(\Gamma_{j})-n_{-}(\Gamma_{j})\Bigr) \ \ \ ({\rm mod}\ 4)\nonumber \\
 =&-\sum_{\Gamma_{j}:{\rm TRIM}}n_{-}(\Gamma_{j})\ \ \ ({\rm mod}\ 4),
\end{align}
where $n_{+}(\Gamma_{j})$ is the number of occupied states with even parity at the TRIM $\Gamma_{j}$.
Therefore,
for systems with inversion symmetry, topological phases are characterized by the symmetry indicator $X_{\rm BS}=(\nu_{1}, \nu_{2}, \nu_{3}, \mu_{1}$) with $\nu_{a}=0, 1$ and $\mu_{1}=0, 1, 2, 3$.

%cuttingprocedure
In Sec.~\ref{cuttingprocedure}, we will show that the gapless hinge states appear in a three-dimensional insulator when $(\nu_{1}, \nu_{2}, \nu_{3}, \mu_{1})=(0,0,0,2)$.
Here, for that purpose, let us consider the numbers of occupied states with odd parity at each TRIM in the case of $(\nu_{1}, \nu_{2}, \nu_{3}, \mu_{1})=(0,0,0,2)$.
As shown in Fig.~\ref{zfourparitypicture}(a), one of the simplest examples to realize $(\nu_{1}, \nu_{2}, \nu_{3}, \mu_{1})=(0,0,0,2)$ is $n_{-}(\Gamma)=2$, $n_{-}(\Gamma_{j})=0$ ($\Gamma_{j}\neq \Gamma$), where $\Gamma=(0, 0, 0)$.
Another example shown in Fig.~\ref{zfourparitypicture}(b) also shows the same set of topological invariants $(\nu_{1}, \nu_{2}, \nu_{3}, \mu_{1})=(0,0,0,2)$,
but this case can be reduced to the case of Fig.~\ref{zfourparitypicture}(a) by gauge transformation of switching the inversion center. 
For example, by shifting the inversion center $\bm{R}$ of the system to $\bm{R}+\bm{a}_{i}/2\ (i=1, 2, 3)$, where $\boldsymbol{a}_{i}$ are translation vectors, the parity at the TRIM $\Gamma_{j}$ on a plane $n_{i}=1$ is multiplied by $(-1)$. Therefore, if we shift the inversion center $\bm{R}$ to $\bm{R}+\bm{a}_{3}/2$, Fig.~\ref{zfourparitypicture}(b) is transformed to (a).
Therefore, Fig.~\ref{zfourparitypicture}(b) is equivalent to Fig.~\ref{zfourparitypicture}(a).
Although there are many cases of patterns of odd-parity states at TRIM with $(\nu_{1}, \nu_{2}, \nu_{3}, \mu_{1})=(0,0,0,2)$ other than Fig.~\ref{zfourparitypicture}(a) and (b),
we will consider the case of Fig.~\ref{zfourparitypicture}(a) for simplicity in this section.
However, a cutting procedure, which is to be discussed in the next subsection, can be applied to every case with  $(\nu_{1}, \nu_{2}, \nu_{3}, \mu_{1})=(0,0,0,2)$ (see Appendix~\ref{section:extension to the general cases}).
%%%%
\subsection{Cutting procedure}
\begin{figure}
\includegraphics{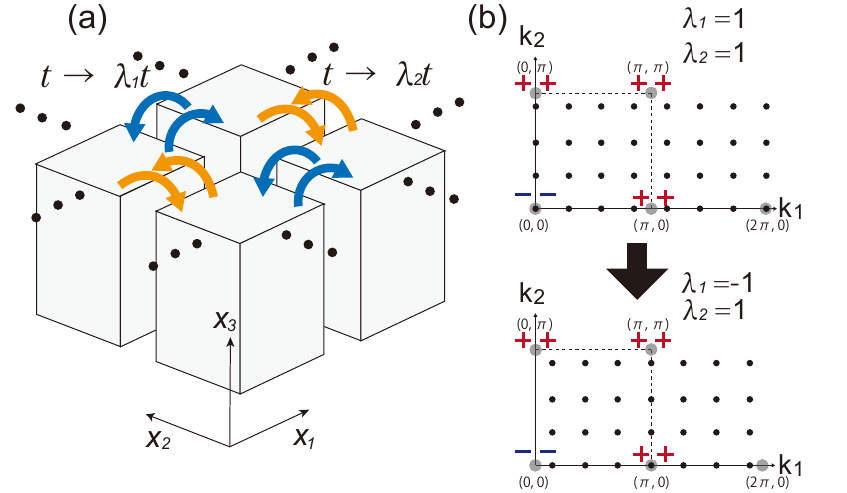}
\caption{\label{cuttingprocedurefig}(Color online) Cutting procedure.~(a) Our boundary conditions for the cutting procedure.~Each gray box represents the entire system, and we show the copies of the system in the figure to illustrate the boundary condition.~Boundary conditions change along the $x_{1}$ and $x_{2}$ directions by changing $\lambda_{1}$ and $\lambda_{2}$.~We replace every hopping amplitude $t$ for the bonds that cross the boundary at $x_{i}=$ const by $\lambda_{i}t$ $(i=1, 2)$.~(b) Black points represent possible wave vectors in the cases of periodic and anti-periodic boundary conditions in the $x_{1}$ direction. Here $\lambda_{2}$ is set to be unity.
When $\lambda_{1}=1$ and $\lambda_{2}=1$, among the four TRIM, only $\bm{k}=(0,0)$ is among the possible wave vectors. Likewise, when $\lambda_{1}=-1$ and $\lambda_{2}=1$, only $\bm{k}=(\pi, 0)$ is among them.
}
\end{figure}
In order to understand the relationship between the $\mathbb{Z}_{4}$ topological index and the hinge states, we introduce a cutting procedure, which is used in the appendix of Ref.~[\onlinecite{PhysRevB.78.045426}] in the context of the time-reversal invariant $\mathbb{Z}_{2}$ topological insulators.
Here, we consider the system to be large but finite, with periodic boundary conditions in the $x_{1}$ and $x_{2}$ directions that are parallel to the primitive reciprocal lattice vectors $\bm{b}_{1}$ and $\bm{b}_{2}$, respectively.
We set the system size as $L_{1}\times L_{2},\ L_{1}=L_{2}=2M+1$ ($M$ is an integer) for simplicity. We will discuss the case with an even number of the system size in Appendix~\ref{section:appendixd}.
Here, the length of the system along the $x_{j}$ direction ($j=1,2,3$) is measured in the unit of the lattice constant, and each unit cell is inversion-symmetric with its inversion center at $x_{j}={\rm integer}$.  
Along the $x_{3}$ direction that is parallel to $\bm{b}_{3}$, we set the periodic boundary condition and set the system size as $L_{3}\rightarrow \infty$.
Thereby, the Bloch wave-vector $k_{3}$ in the $x_{3}$ direction can be defined. 
In this subsection, we focus on $k_{3}=0$. That is, we will consider the four TRIM $\Gamma_{j=(n_{1},n_{2},0)}$ shown in Fig.~\ref{zfourparitypicture}(a).

Along the $x_{1}$ and $x_{2}$ directions, instead of the periodic boundary conditions, we multiply all the hopping amplitudes across the boundary between $x_{i}=-M$ and $x_{i}=M$ by a real parameter $\lambda_{i}$. 
This means that the boundary conditions for the finite systems in the $x_{1}, x_{2}$ directions change by changing $\lambda_{1},\ \lambda_{2}$, as shown in Fig.~\ref{cuttingprocedurefig}(a). The case with $\lambda_{1}=1$ corresponds to periodic boundary condition in the $x_{1}$ direction and that with $\lambda_{1}=0$ corresponds to an open boundary condition in the $x_{1}$ direction.
 For any values of $\lambda_{1}$ and $\lambda_{2}$, the system is inversion symmetric with its inversion center at $(x_{1},x_{2},x_{3})=(0,0,0)$.
 
\subsection{Spectral flows in the band gap\label{cuttingprocedure}}
In the following, we show existence of gapless hinge states when $(\nu_{1},\nu_{2},\nu_{3},\mu_{1})=(0,0,0,2)$, i.e. the $\mathbb{Z}_{4}$ index is nontrivial. In the cutting procedure with the parameters $\lambda_{1}$ and $\lambda_{2}$, the hinge states appear at $\lambda_{1}=\lambda_{2}=0$, while the bulk topological invariants $(\nu_{1},\nu_{2},\nu_{3},\mu_{1})$ determine the parity eigenvalues of the states at $(\lambda_{1},\lambda_{2})=(1,\pm1)$, $(-1,\pm1)$ as we show in the following. To relate the information of wave-functions at $(\lambda_{1},\lambda_{2})=(1,\pm1)$, $(-1,\pm1)$ with that at $\lambda_{1}=\lambda_{2}=0$, we utilize symmetry of the spectral flows under $\lambda_{1}\leftrightarrow -\lambda_{1}$ and under $\lambda_{2}\leftrightarrow -\lambda_{2}$, as long as the surface is gapped.

When we consider the case with $\lambda_{1}=1$,
the 
wave-vector in the $x_{1}$ direction is
\begin{equation}
k_{1}=\frac{2\pi}{L_{1}}m_{1}\ \ \ (-M\leq m_{1}\leq M),
\end{equation}
because of the periodic boundary condition in the $x_{1}$ direction.
Because $L_{1}$ is an odd number, $k_{1}$ can be $0$ but not $\pi$.
When $(\lambda_{1},\lambda_{2})=(1,1)$,
$(k_{1}, k_{2})$ can take a value $(0, 0)$, but not
 $(\pi, 0),\ (0, \pi)$ or $(\pi, \pi)$ as shown in Fig.~\ref{cuttingprocedurefig}(b).
 
 \begin{figure}
\includegraphics{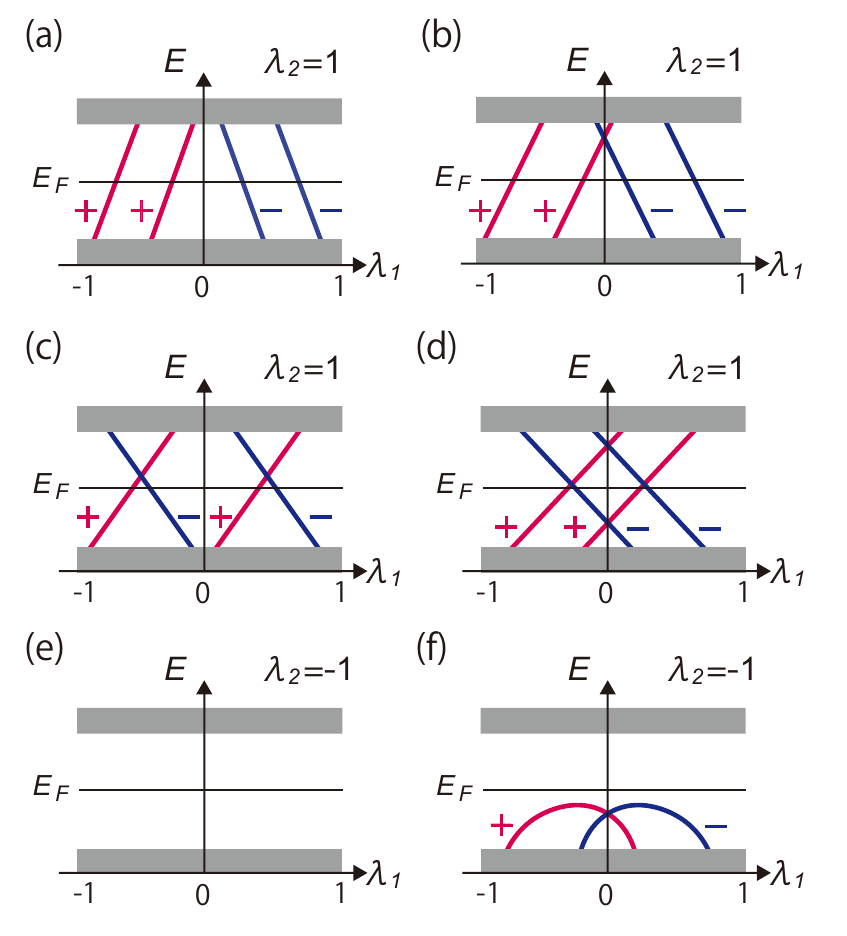}
\caption{\label{energylambda}(Color online) Energy spectra in changing $\lambda_{1}=1\ \rightarrow -1$ with $\lambda_{2}$ being constant. The energy spectra are symmetric with respect to $\lambda_{1}\leftrightarrow -\lambda_{1}$, and states at $\lambda_{1}$ and $-\lambda_{1}$ have opposite parity eigenvalues. (a-d) are four representative examples when $\lambda_{2}=1$. (e) and (f) are two examples when $\lambda_{2}=-1$.}
\end{figure}

Next, we consider the case with $\lambda_{1}=-1$. In this case, wave functions are multiplied by $-1$ across the boundary between $x_{1}=M$ and $x_{1}=-M$, corresponding to an anti-periodic boundary condition in the $x_{1}$ direction.
This anti-periodic boundary condition is converted into the periodic boundary condition in the $x_{1}$ direction by a unitary transformation $U_{1}={\rm exp}[i\pi \hat{x}_{1}/L_{1}]$, where $\hat{x}_{1}$ is a position operator for the coordinate $x_{1}$. Through this transformation, the Bloch wave vector is shifted as $k_{1} \rightarrow k_{1}+\frac{\pi}{L_{1}}$ due to this unitary transformation $U_{1}$ (see Appendix \ref{apendixa}).
Thus, the Bloch wave vector in the $x_{1}$ direction is
\begin{equation}
k_{1}=\frac{2\pi}{L_{1}}m_{1}+\frac{\pi}{L_{1}}\ \ \ (-M\leq m_{1}\leq M).
\end{equation}
In this case, because $L_{1}$ is an odd number, $k_{1}$ can be $\pi$ but not $0$.
When $(\lambda_{1},\lambda_{2})=(-1,1)$,
$(k_{1}, k_{2})$ can take a value $(\pi, 0)$, but not
 $(0, 0),\ (0, \pi),\ (\pi, \pi)$ as shown in Fig.~\ref{cuttingprocedurefig}(b).
 
Now, we calculate $N_{-}(\lambda_{1},\lambda_{2})$, representing the number of occupied states with odd parity at $k_{3}=0$.
Inversion operation $\hat{I}$ changes $(k_{1}, k_{2}, k_{3})$ to $(-k_{1}, -k_{2}, -k_{3})$. 
Each wave-function $\psi_{m}(\boldsymbol{k})$ at non-TRIM points $\boldsymbol{k}=(k_{1},k_{2},k_{3})$ with $k_{3}=0$ can always be paired with one at $-\boldsymbol{k}$ to construct two states, one with even-parity $\phi_{+}$ and the other with odd-parity $\phi_{-}$: 
\begin{equation}
\phi_{\pm}\equiv \frac{1}{\sqrt{2}}\biggl(\psi_{m}(\boldsymbol{k})\pm \hat{I}\psi_{m}(\boldsymbol{k})\biggr),
\end{equation}
where $\hat{I}\psi_{m}(\boldsymbol{k})\propto \psi_{m}(-\boldsymbol{k})$.
Therefore, each non-TRIM pair ($\boldsymbol{k}$, $-\boldsymbol{k}$) with $k_{3}=0$ contributes 1 to $N_{-}(\lambda_{1},\lambda_{2})$.

On the other hand, a contribution to $N_{-}(\lambda_{1},\lambda_{2})$ from the TRIM depends on $\lambda_{1}$ and $\lambda_{2}$. First, we consider the case with $\lambda_{2}=1$.
When $(\lambda_{1},\lambda_{2})=(1, 1)$, the number of odd-parity states at TRIM that contributes to $N_{-}(\lambda_{1},\lambda_{2})$ is
$n_{-}(0,0,0)$, where $n_{-}(k_{1}, k_{2}, k_{3})$ is the number of occupied states with odd parity at TRIM $\Gamma_{j=(n_{1},n_{2},n_{3})}=(n_{1}\boldsymbol{b}_{1}+n_{2}\boldsymbol{b}_{2}+n_{3}\boldsymbol{b}_{3})/2$.
Let $\nu$ be the number of occupied bands. Then $N_{-}(\lambda_{1}=1, \lambda_{2}=1)$ can be expressed as follows:
\begin{equation}\label{noddlambda1}
N_{-}(1, 1)=\frac{(L_{1}L_{2}-1)\nu}{2}+n_{-}(0,0,0).
\end{equation}
Similarly, when $(\lambda_{1}, \lambda_{2})=(-1, 1)$, among the  TRIM only the TRIM $\Gamma_{j=(1,0,0)}=(\pi,0,0)$ contributes to $N_{-}(\lambda_{1},\lambda_{2})$.
From this, $N_{-}(\lambda_{1}=-1,\lambda_{2}=1)$ can be expressed as follows:
\begin{equation}\label{noddlambdaminus1}
N_{-}(-1, 1)=\frac{(L_{1}L_{2}-1)\nu}{2}+n_{-}(\pi,0,0).
\end{equation}
Therefore, 
from Eqs.~(\ref{noddlambda1}) and (\ref{noddlambdaminus1}), the total change in $N_{-}(\lambda_{1}, \lambda_{2}=1)$ between $\lambda_{1}=1$ and $\lambda_{1}=-1$ can be expressed as follows:
\begin{align}\label{kyusiki}
&\bigl[N_{-}(\lambda_{1}, \lambda_{2}=1)\bigr]^{\lambda_{1}=1}_{\lambda_{1}=-1}
\nonumber \\
=& n_{-}(0,0,0)-n_{-}(\pi,0,0)=2,
\end{align}
where
\begin{align}
\bigl[N_{\pm}(\lambda_{1}, \lambda_{2})\bigr]^{\lambda_{1}=a}_{\lambda_{1}=b}\equiv N_{\pm}(a,\lambda_{2})
- N_{\pm}(b,\lambda_{2}).
\end{align}
That is, in the process of changing from $\lambda_{1}=1$ to $-1$,
the number of occupied states with odd parity is reduced by 2. 
In addition, we can show that the energy spectrum is symmetric with respect to $\lambda_{1}\leftrightarrow -\lambda_{1}$, and the bound states $\ket{\psi_{l}(\lambda_{1})}$ and $\ket{\psi_{l}(-\lambda_{1})}$ have opposite parity eigenvalues (see Appendix \ref{apendixa}).
Thus, as we show some examples in Fig.~\ref{energylambda}(a-d), two states with odd-parity move from the valence bands for $\lambda_{1}=1$ to the conduction bands for $\lambda_{1}=-1$. 
In addition, two states with even-parity move from the conduction bands for $\lambda_{1}=1$ to the valence bands for $\lambda_{1}=-1$. 
This means that the following relation generally holds:
\begin{equation}\label{atarasikigou}
\bigl[N_{\pm}(\lambda_{1},\lambda_{2}=1)\bigr]^{\lambda_{1}=0}_{\lambda_{1}=1}=\bigl[N_{\mp}(\lambda_{1}, \lambda_{2}=1)\bigr]^{\lambda_{1}=0}_{\lambda_{1}=-1}.
\end{equation}

Now, we calculate the $N_{+-}(\lambda_{1},\lambda_{2})\equiv N_{+}(\lambda_{1},\lambda_{2})-N_{-}(\lambda_{1},\lambda_{2})$ for the four points A  $(\lambda_{1}=1,\lambda_{2}=1)$, B $(\lambda_{1}=0,\lambda_{2}=1)$, C $(\lambda_{1}=1,\lambda_{2}=-1)$ and D $(\lambda_{1}=0, \lambda_{2}=-1)$ as shown in Fig.~\ref{fig:appearing_hinge}(a). We calculate the differences of this value between (i) A-B, (ii) C-D and (iii) A-C in the followings.

(i) First, we calculate the change in $N_{+-}(\lambda_{1},\lambda_{2})$ between the point A and the point B.
From Eqs.~(\ref{kyusiki}) and (\ref{atarasikigou}), we obtain the following relation.
\begin{align}\label{hukusen_appearance_ichi}
&\bigl[N_{+-}(\lambda_{1}, \lambda_{2}=1)\bigr]^{\lambda_{1}=0}_{\lambda_{1}=1}\nonumber \\
=&\bigl[N_{-}(\lambda_{1}, \lambda_{2}=1)\bigr]^{\lambda_{1}=0}_{\lambda_{1}=-1}-
\bigl[N_{-}(\lambda_{1}, \lambda_{2}=1)\bigr]^{\lambda_{1}=0}_{\lambda_{1}=1}\nonumber \\
=&\bigl[N_{-}(\lambda_{1}, \lambda_{2}=1)\bigr]^{\lambda_{1}=1}_{\lambda_{1}=-1}=2.
\end{align}

(ii) Next, we analyze the case with $\lambda_{2}=-1$. 
In this case, 
$N_{-}(\lambda_{1}, \lambda_{2}=-1)$ can be expressed as follows:
\begin{equation}
N_{-}(1, -1)=\frac{(L_{1}L_{2}-1)\nu}{2}+n_{-}(0,\pi,0),
\end{equation}
\begin{equation}
N_{-}(-1, -1)=\frac{(L_{1}L_{2}-1)\nu}{2}+n_{-}(\pi,\pi,0),
\end{equation}
in the same way as in the case of $\lambda_{2}=1$.
Therefore,
the total change in $N_{-}(\lambda_{1}, \lambda_{2}=-1)$ between the point C and the point E ($\lambda_{1}=-1, \lambda_{2}=-1$), can be expressed as follows:
\begin{align}\label{minusichinikoteisiteichikaraminuichi}
&\bigl[N_{-}(\lambda_{1}, \lambda_{2}=-1)\bigr]^{\lambda_{1}=1}_{\lambda_{1}=-1}\nonumber \\
=&n_{-}(0,\pi,0)-n_{-}(\pi,\pi,0)=0.
\end{align}
That is,
the number of occupied states with odd parity does not change
in the process of changing from $\lambda_{1}=1$ to $-1$.
Thus, 
the change of the energy spectra are as shown in Fig.~\ref{energylambda}(e)(f)
so as to satisfy
$N_{-}(1, -1)=N_{-}(-1, -1)$
because states with even parity and odd parity are mutually transformed by $\lambda_{1}\leftrightarrow -\lambda_{1}$.
 From Eqs.~(\ref{atarasikigou}) and (\ref{minusichinikoteisiteichikaraminuichi}), the change in $N_{+-}(\lambda_{1},\lambda_{2}=-1)$ between the point C and the point D, can be expressed as follows:
\begin{align}\label{N(0,-1)nosa}
&\bigl[N_{+-}(\lambda_{1}, \lambda_{2}=-1)\bigr]^{\lambda_{1}=0}_{\lambda_{1}=1}\nonumber \\
=&\bigl[N_{-}(\lambda_{1}, \lambda_{2}=-1)\bigr]^{\lambda_{1}=0}_{\lambda_{1}=-1}
-\bigl[N_{-}(\lambda_{1}, \lambda_{2}=-1)\bigr]^{\lambda_{1}=0}_{\lambda_{1}=1}\nonumber \\
=&\bigl[N_{-}(\lambda_{1}, \lambda_{2}=-1)\bigr]^{\lambda_{1}=1}_{\lambda_{1}=-1}=0.
\end{align}

(iii) In the above, we have considered the energy spectra when we fix $\lambda_{2}=1$ or $\lambda_{2}=-1$ and vary $\lambda_{1}$.
Similarly, we also get the similar conclusion when we fix $\lambda_{1}=1$ or $\lambda_{1}=-1$ and  vary $\lambda_{2}$.
Therefore, similarly to Eq.~(\ref{kyusiki}), the total change in $N_{-}(\lambda_{1}=1,\lambda_{2})$ between the point A and the point C in Fig.~\ref{fig:appearing_hinge}(a), can be expressed as follows:
\begin{align}
&\bigl[ N_{-}(\lambda_{1}=1,\lambda_{2}) \bigr]^{\lambda_{2}=1}_{\lambda_{2}=-1}\nonumber \\
=&n_{-}(0,0,0)-n_{-}(0,\pi,0)=2.
\end{align}
That is, two states with odd parity move from the valence bands of $\lambda_{2}=1$ to the conduction bands of $\lambda_{2}=-1$.
In the same way as when $\lambda_{1}$ varies, 
we can show that the energy spectra are symmetric with respect to $\lambda_{2} \leftrightarrow -\lambda_{2}$, and the bound states $\ket{\psi(\lambda_{2})}$ and $\ket{\psi(-\lambda_{2})}$ have opposite parity.
Therefore, two states with even-parity move from the conduction bands of $\lambda_{2}=1$ to the valence bands of $\lambda_{2}=-1$.
From this discussion, we conclude that the number $N_{+-}(\lambda_{1},\lambda_{2})$ increases by 4 by changing from the point A to the point C.
This means that the following relation holds:
\begin{align}\label{N(1,-1)nosa}
\bigl[ N_{+-}(\lambda_{1}=1,\lambda_{2})\bigl]^{\lambda_{2}=-1}_{\lambda_{2}=1}=4.
\end{align}

To summarize (i)-(iii),
from Eqs.~(\ref{hukusen_appearance_ichi}), (\ref{N(0,-1)nosa}) and (\ref{N(1,-1)nosa}), we obtain the following equation.
\begin{align}\label{mainresult2b}
&\bigl[N_{+-}(\lambda_{1}=0,\lambda_{2}) \bigr]^{\lambda_{2}=-1}_{\lambda_{2}=1}=2.
\end{align}
This equation is the main result of this subsection and this is closely related to the appearance of   hinge states as described in the next subsection.
In the next subsection, we will consider the energy spectrum when $\lambda_{1}=0$ through the change of $\lambda_{2}$ from $\lambda_{2}=1$ to $\lambda_{2}=-1$, and for this purpose Eq.~(\ref{mainresult2b}) is important.

%%subsection
\subsection{Appearance of hinge states}\label{subsec:Appearance of hinge states}
%fig
\begin{figure}
\includegraphics{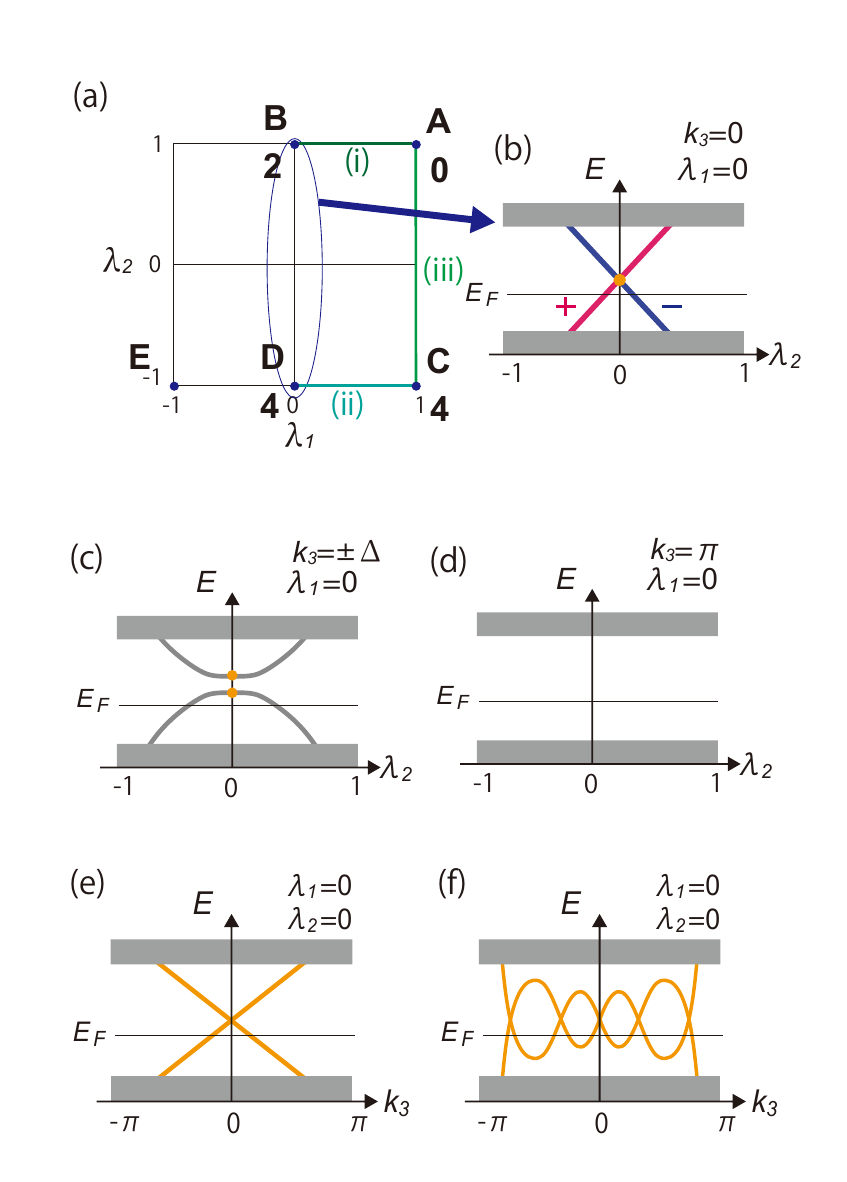}
\caption{\label{fig:appearing_hinge}(Color online) Appearance of the hinge states. (a) The differences in $N_{+-} (\lambda_{1},\lambda_{2})\equiv N_{+}(\lambda_{1},\lambda_{2})-N_{-}(\lambda_{1},\lambda_{2})$ at the points A, B, C and D with respect to the point A for $k_{3}=0$. From this we conclude that $N_{+-}(\lambda_{1},\lambda_{2})$ at the points B and D are different by two. (b) The energy spectrum when $\lambda_{1}=0$ and $k_{3}=0$ with changing $\lambda_{2}$ from $1$ to $-1$. (c) The energy spectrum when $\lambda_{1}=0$ and $k_{3}=\pm \Delta$, where $\Delta$ represents a small real number. Because $k_{3}$ is away from $k_{3}=0$, degeneracy between the even-parity and the odd-parity states at $k_{3}=0$ is lifted because of absence of inversion symmetry for $k_{3}=\pm \Delta$. (d) The energy spectrum when $\lambda_{1}=0$ and $k_{3}=\pi$. In this case, states with even and odd parity do not cross. 
(e) The band structure when $(\lambda_{1},\lambda_{2})=(0,0)$. 
Two gapless states are degenerate at $k_{3}=0$ corresponding to the yellow point at $\lambda_{2}=0$ in (b). This degeneracy is lifted when $k_{3}\neq 0$ corresponding to the yellow points in (c).
These two states move to conduction bands and valence bands from $k_{3}=\Delta$ to $k_{3}=\pi$, corresponding to (d).
(f) An example of the band structure crossing the Fermi level in a more complicated manner than (e).
 }
\end{figure}
%fig
Here we consider the energy spectrum $k_{3}=0$ when $\lambda_{1}=0$.
From Eq.~(\ref{mainresult2b}), we find that the number $N_{+-}(\lambda_{1},\lambda_{2})\equiv N_{+}(\lambda_{1},\lambda_{2})-N_{-}(\lambda_{1},\lambda_{2})$ increases by 2 when $\lambda_{2}$ is changed from $\lambda_{2}=1$ to $-1$.
Here we note that the energy spectra are symmetric with respect to $\lambda_{2}\leftrightarrow -\lambda_{2}$ even at $\lambda_{1}=0$, and the states $\ket{\psi_{l}(\lambda_{2})}$ and $\ket{\psi_{l}(-\lambda_{2})}$ have opposite parities by assuming that no gapless state appears in the surface.
From this result, we find that in this process of changing $\lambda_{2}$, the exchange of states with odd and even parity takes place once as shown in Fig.~\ref{fig:appearing_hinge}(b).
From this argument, we conclude that even-parity states and odd-parity ones must be degenerate at $(\lambda_{1},\lambda_{2})=(0,0)$ as shown in Fig.~\ref{fig:appearing_hinge}(b).
%fig

In the arguments so far, we considered the case with $k_{3}=0$.
On the other hand, when $k_{3}=\pm \Delta$, where $\Delta$ represents a small non-zero real number, the wave-vector $\boldsymbol{k}$ is not a TRIM. Therefore the states with even and odd parity hybridize, and a gap opens at $\boldsymbol{k}$ when 
$k_{3}=\Delta$. Therefore, the energy spectrum is shown in Fig.~\ref{fig:appearing_hinge}(c) as $\lambda_{2}$ is changed from $\lambda_{2}=1$ to $\lambda_{2}=-1$.
\begin{figure}
\includegraphics{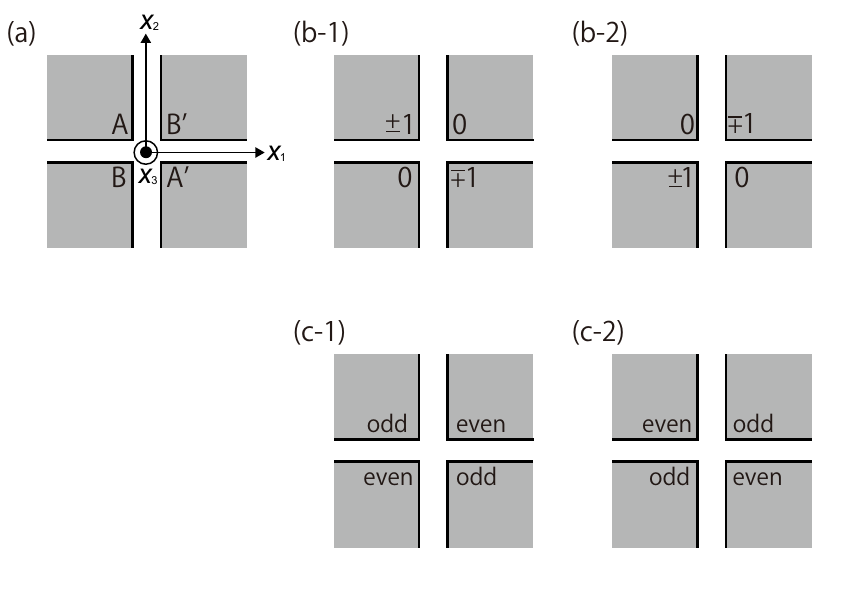}
\caption{\label{tuikahingefig}The numbers of hinge states at four hinges. (a) Four hinges $\rm A, A', B, B'$ produced by the cutting procedure. (b, c) Let $n_{i}$ ($i={\rm A, A', B, B'}$) denote the number of hinge modes at the $i$-hinge. We find that $n_{\rm A}=-n_{\rm A'}$, $n_{\rm B}=-n_{\rm B'}$ and $n_{\rm A}+n_{\rm B}$ is an odd number. One of $n_{\rm A}(=-n_{\rm A'})$ and $n_{\rm B}(=-n_{\rm B'})$ is odd while the other is even. (b-1) and (b-2) are minimal  configurations. (c-1) and (c-2) are general ones.}
\end{figure}

\begin{figure*}
\includegraphics{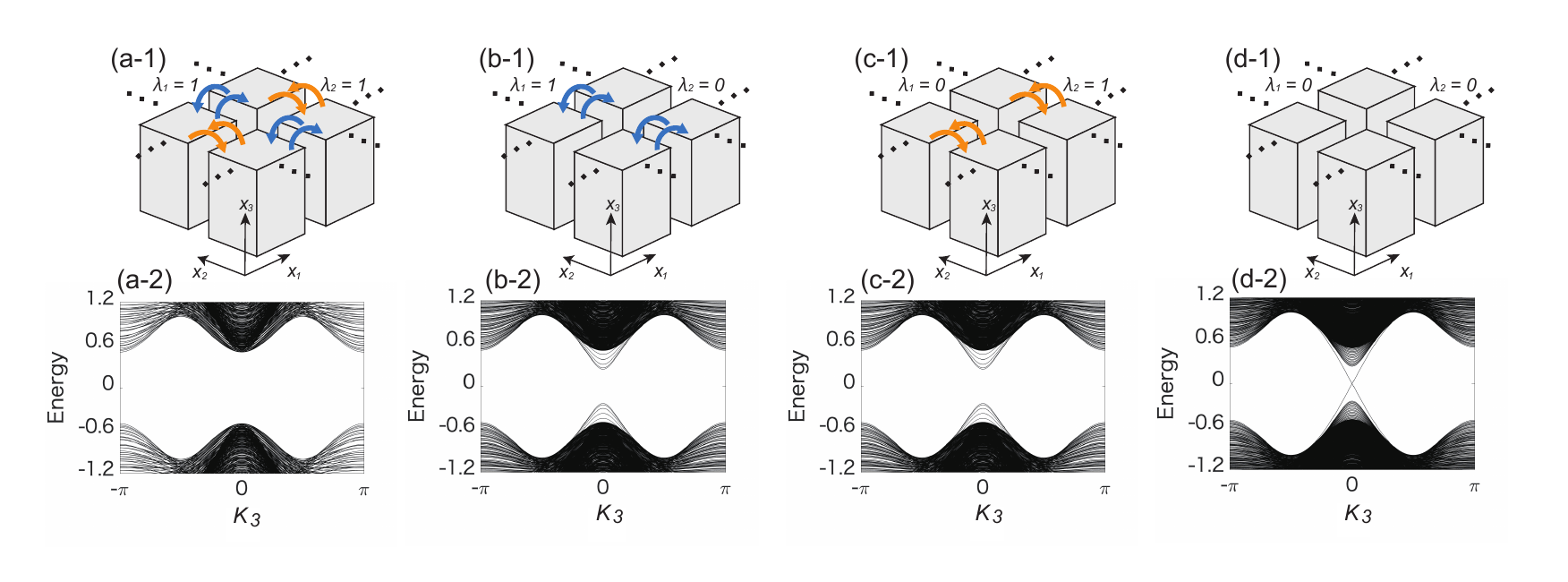}
\caption{\label{systems_4band}Band structures of the tight-binding model (\ref{bhzmodel_ziba})
with parameters $t=c=1$, $m=2$, $B=1/2$ and $\theta=\pi/4$. The boundary condition in the $x_{3}$ direction is periodic and that in the $x_{1}$ and $x_{2}$ directions are characterized by $\lambda_{1}$ and $\lambda_{2}$. Parameter values for boundary conditions are (a) $\lambda_{1}=1$, $\lambda_{2}=1$, (b) $\lambda_{1}=1$, $\lambda_{2}=0$,  (c) $\lambda_{1}=0$, $\lambda_{2}=1$ and (d) $\lambda_{1}=0$, $\lambda_{2}=0$. The system size along $x_{1}$ and $x_{2}$ directions are $L_{1}=L_{2}=45$.
}
\end{figure*}

Next, we consider the spectrum at $k_{3}=\pi$, following the discussion in Sec.~\ref{cuttingprocedure}. In contrast with $k_{3}=0$,
there is no difference in the number of odd-parity eigenstates at the four TRIM on $k_{3}=\pi$.
Then, we can conclude that states with odd parity and states with even parity are not exchanged  on $k_{3}=\pi$ as shown in Fig.~\ref{fig:appearing_hinge}(d).
Namely, the degeneracy of states at $(\lambda_{1},\lambda_{2})=(0,0)$ is present only at $k_{3}=0$ but not at $k_{3}=\pi$.
Therefore gapless states appear when $(\lambda_{1},\lambda_{2})=(0,0)$ as shown in Fig.~\ref{fig:appearing_hinge}(e). 
These gapless states are hinge states because they appear only when there are no bonds across the two boundaries along the $x_{1}$ and $x_{2}$ directions (see Fig.~\ref{cuttingprocedurefig}(a)). Therefore, this system is a SOTI.

From the above discussion, we can conclude the followings for $(\lambda_{1},\lambda_{2})=(0,0)$. 
(i) The two hinge states are degenerated at $k_{3}=0$. (ii) By changing from $k_{3}=0$ to $k_{3}=\pi$, one of the degenerated states moves to the valence band and the other moves to the conduction band. 
Then a band structure of the hinge states of the SOTI can be like Fig.~\ref{fig:appearing_hinge}(e).
We note that Fig.~\ref{fig:appearing_hinge}(e) is only an example, and the band structure can be different from Fig.~\ref{fig:appearing_hinge}(e).
In such cases, we can conclude from (i) and (ii) that the number of states crossing the Fermi level between $k_{3}=0$ and $k_{3}=\pi$ in the band structure of the SOTI will always be an odd number (for example see Fig.~\ref{fig:appearing_hinge}(f)).
Note that we consider the case of Fig.~\ref{zfourparitypicture}(a) as parity eigenvalues at TRIM in the arguments so far. We extend the discussion so far to general cases of parity eigenvalues at TRIM in Appendix~\ref{section:extension to the general cases}.
In addition, while we find that a pair of hinge states with positive and negative velocities appears as a particular example in this section,  we conclude that in general an odd number of the pairs of hinge states always appear when $(\nu_1,\nu_2,\nu_3,\mu_1)=(0,0,0,2)$ from the discussion in  Appendix.~\ref{section:extension to the general cases}.

When $\lambda_{1}=\lambda_{2}=0$, the cutting produces four hinges, and the hinge states exist at one of the four hinges. If one hinge state lies on the hinge A with a positive velocity along $x_{3}$ in Fig.~\ref{tuikahingefig}(a) as an example, the inversion symmetry imposes that the hinge $\rm A'$ supports a hinge states with a negative velocity along $x_{3}$. It is also true for the pair of hinges B and $\rm B'$. From these considerations, we conclude that the hinges A and $\rm A'$ (and likewise B and $\rm B'$) have the same number of hinge states, and their hinge states form pairs under inversion symmetry, i.e. having opposite signs of velocities. Let $n_{i}$ ($i={\rm A},{\rm A'}, {\rm B}, {\rm B'}$) denote the number of hinge modes at the $i$-hinge. We define this number to be the number of hinge modes with positive velocity minus that with negative velocity. Then from the above argument, we conclude that $n_{\rm A}=-n_{\rm A'}$, $n_{\rm B}=-n_{\rm B'}$ and $n_{\rm A}+n_{\rm B}$ is an odd number. Then, minimal configurations for $n_{i}$ are shown in Figs.~\ref{tuikahingefig}(b-1) and (b-2). In general, we conclude that one of $n_{\rm A}(=-n_{\rm A'})$ and $n_{\rm B}(=-n_{\rm B'})$ is odd while the other is even as shown in Figs.~\ref{tuikahingefig}(c-1) and (c-2).

So far we showed that $(\nu_{1},\nu_{2},\nu_{3},\mu_{1})=(0,0,0,2)$ leads to existence of hinge states. Here we explain  that in a case with $(\nu_{1},\nu_{2},\nu_{3})\neq (0,0,0)$ i.e. a three-dimensional Chern insulator, existence of hinge states does not follow.
For example, when $(\nu_{1},\nu_{2},\nu_{3})=(0,1,0)$, the system is a Chern insulator, and there exist chiral surface states on the $x_{2}$-$x_{3}$ surface.
In this case, along the $\lambda_{1}=0$ line in Fig.~\ref{fig:appearing_hinge}(a), spectral symmetry under $\lambda_{2}\leftrightarrow -\lambda_{2}$ does not hold, because existence of the gapless surface states invalidates the proof for the $\lambda_{2}\leftrightarrow -\lambda_{2}$ symmetry in Appendix \ref{appendixa1proofofeforlofalized}. It physically means that the chiral surface states hide hinge states if any. Thus, to summarize, when $(\nu_{1},\nu_{2},\nu_{3})\neq(0,0,0)$, $\mu_{1}=2$ does not lead to the existence of hinge states.
%%section3
\section{\label{section:Tight-binding model}Tight-binding model}
%fig
Here, we perform a model calculation to verify the arguments in the previous section.
We start from a tight-binding model of a SOTI on a simple-cubic lattice with inversion symmetry [\onlinecite{PhysRevB.98.205129}].
In order to construct the model of the SOTI, we add a uniform Zeeman magnetic field $\boldsymbol{B}=B(-\sin{\theta}, \cos{\theta},0)$ to the model of the three-dimensional TI in Ref.~[\onlinecite{PhysRevB.78.195424}].
By adding a Zeeman term $-\tau_{0}\otimes \boldsymbol{B}\cdot \boldsymbol{\sigma}$ to the model of the TI,
we get a four-band tight-binding model given by
\begin{align}\label{bhzmodel_ziba}
H(\boldsymbol{k})=&-t\sum_{j}\sin{k_{j}}\tau_{1}\otimes \sigma_{j}\nonumber \\
&-(m-c\sum_{j}\cos{k_{j}})\tau_{3}\otimes \sigma_{0}
-\tau_{0}\otimes \boldsymbol{B}\cdot \boldsymbol{\sigma},
\end{align}
where $\tau_{j}$ and $\sigma_{j}\ (j=1, 2, 3)$ are Pauli matrices, and $\tau_{0}$ and $\sigma_{0}$ are the $2\times 2$ identity matrices.
In this model,
the term $-\tau_{0}\otimes \boldsymbol{B}\cdot \boldsymbol{\sigma}$ breaks symmetry under the time-reversal operator $T=-i\tau_{0}\otimes \sigma_{2}$ but respects symmetry under the inversion operator $I=\tau_{3}\otimes \sigma_{0}$.
To realize a SOTI phase,
we set $t=c=1,\ m=2,\ B=1/2$ and $\theta=\pi/4$ in the following.
We set the Fermi energy to be $E_{F}=0$.

\begin{figure}
\includegraphics{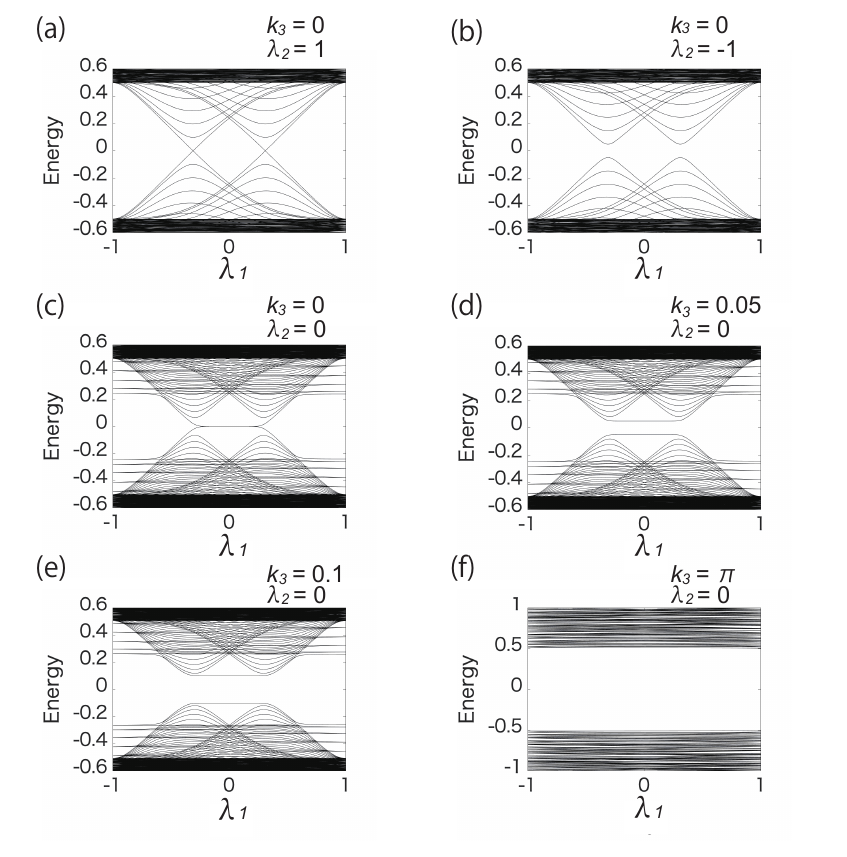}
\caption{\label{fig:lamz_lamy}Energy spectra with changing $\lambda_{1}$ from $\lambda_{1}=-1$ to $1$
for the model (\ref{bhzmodel_ziba}).
Parameters are set as $t=c=1$, $m=2$, $B=1/2$ and $\theta=\pi/4$.
(a) $\lambda_{2}=1$ and $k_{3}=0$. (b) $\lambda_{2}=-1$ and $k_{3}=0$. (c) $\lambda_{2}=0$ and $k_{3}=0$. (d) $\lambda_{2}=0$ and $k_{3}=0.05$. (e) $\lambda_{2}=0$ and $k_{3}=0.1$. (f) $\lambda_{2}=0$ and $k_{3}=\pi$.}
\end{figure}

In this model, surface Dirac cones perpendicular to either $x_{1}$ and $x_{2}$ axes are gapped by the uniform magnetic field. 
Between the surfaces with an inward magnetic field and those with an outward magnetic field, the signs of the mass term of the surface Dirac cones are opposite. Therefore, at the intersections of these two surfaces, gapless states necessarily appear. 
In this model,
only the $\Gamma$ point [$\boldsymbol{k}=(0, 0, 0)$] has two odd- parity states and other TRIM have only even-parity states as in Fig.~\ref{zfourparitypicture}(a).
From these parity eigenvalues,
we get the weak indices $\nu_{1}=\nu_{2}=\nu_{3}=0$, and the strong index $\mu_{1}=2$.

Here we set a periodic boundary condition in the $x_{3}$ direction.
The system has a size $L_{1}\times L_{2}$ along $x_{1}$ and $x_{2}$ directions, and we first set periodic boundary conditions along the $x_{1}$ and $x_{2}$ directions with $L=2M+1$.
In the calculation we set the system size as $L\times L=45\times 45$. 
We then replace the hopping amplitudes $t$ and $c$ for all bonds across the boundary between $x_{1}=-M$ and $x_{1}=M$ by $\lambda_{1}t$ and $\lambda_{1}c$, where $\lambda_{1}$ is real. In addition we similarly replace the hopping amplitudes for all bonds that cross the boundary between $x_{2}=-M$ and $x_{2}=M$ by $\lambda_{2}t$ and $\lambda_{2}c$, where $\lambda_{2}$ is real. 

%%fig
\begin{figure*}
\includegraphics{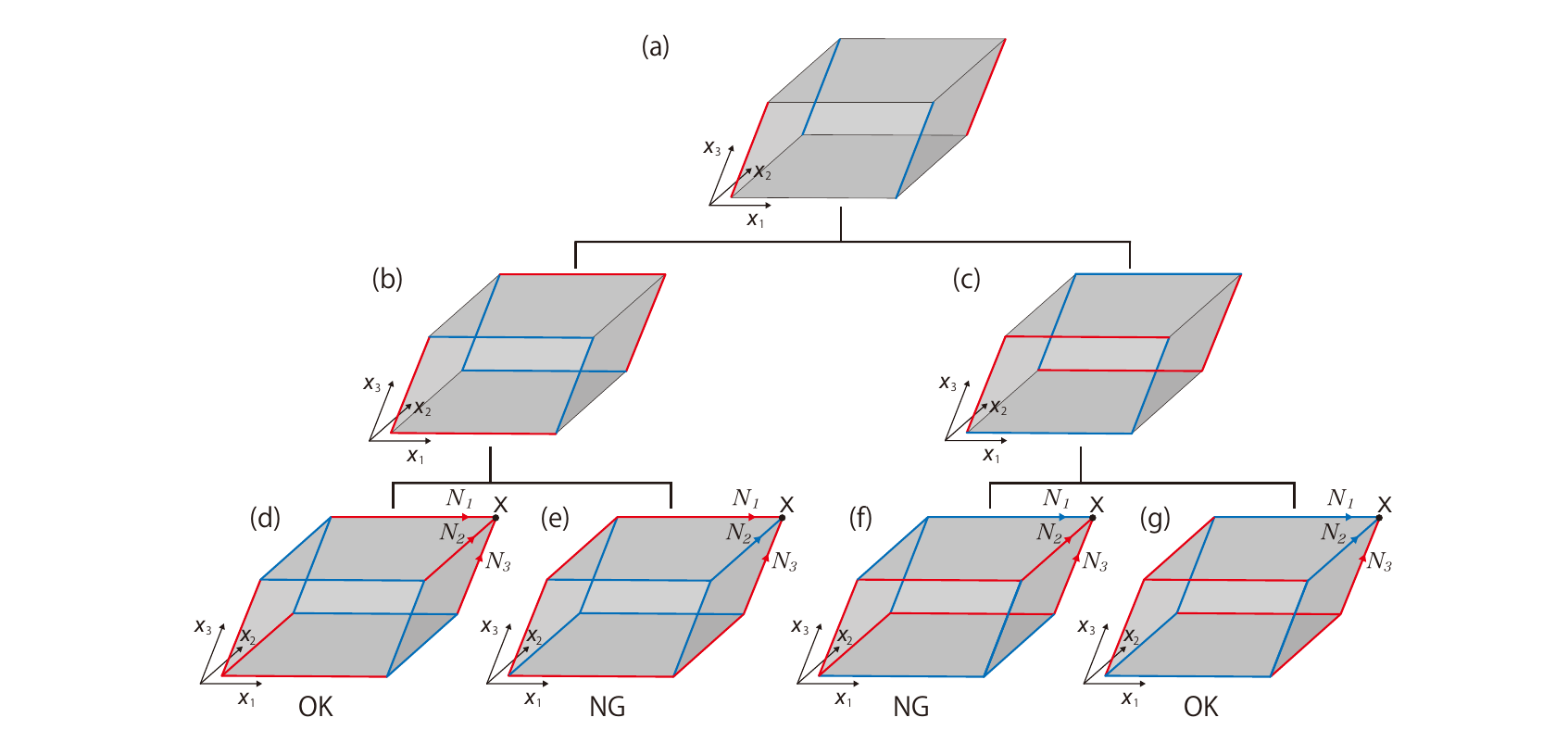}% Here is how to import EPS art
\caption{\label{fig:location_hinge}(Color online)
Positions of hinge states. Blue lines represent an odd number of pairs of hinge states and red lines represent an even number of pairs of hinge states.
(a) Hinge states appear along $x_{3}$ directions.
(b, c) By introducing cutting procedure along $x_{2}$ and $x_{3}$ directions, it is found that hinge states appear along $x_{1}$ direction.
(d-g) Hinge states along $x_{1}$, $x_{2}$ and $x_{3}$ directions. The number of hinge modes toward the corner $X$ of the system are defined as $N_{1}$, $N_{2}$ and $N_{3}$ respectively. 
Realizable positions of hinge modes can only be (d) and (g) because  of charge conservation at the corner $X$.}
\end{figure*}

First we calculate the band structures of the model when ($\lambda_{1},\lambda_{2}$)=($1,1$), ($1,0$), ($0,1$) and ($0,0$). The results are shown in Figs.~\ref{systems_4band}(a-2), (b-2), (c-2) and (d-2) respectively. In addition the schematic figures corresponding to these results are Figs.~\ref{systems_4band}(a-1), (b-1), (c-1) and (d-1). From these results, we find that gapless states appear only in Fig.~\ref{systems_4band}(d-2).
In Fig.~\ref{systems_4band}(a) ($\lambda_{1}=\lambda_{2}=1$), the system has no boundary, and the eigenstates are bulk states and are gapped. In Fig.~\ref{systems_4band}(b) ($\lambda_{1}=1$, $\lambda_{2}=0$) and (c) ($\lambda_{1}=0$, $\lambda_{2}=1$) the system has surfaces, and the results in (b-2) and (c-2) show that the surface spectrum is also gapped. In Fig.~\ref{systems_4band}(d) ($\lambda_{1}=\lambda_{2}=0$), the system has surfaces and hinges. Therefore, the gapless states in Fig.~\ref{systems_4band}(d-2) are hinge states.

Next we calculate a change in the energy spectra from $\lambda_{1}=-1$ to $1$ when $k_{3}$ and $\lambda_{2}$ are fixed. The results are shown in Fig.~\ref{fig:lamz_lamy}. 
When $\lambda_{1}$ changes from $\lambda_{1}=-1$ to $\lambda_{1}=1$, two states are interchanged between the conduction and the valence bands when $\lambda_{2}=1$ and $k_{3}=0$ (Fig.~\ref{fig:lamz_lamy}(a)) but not when $\lambda_{2}=-1$ and $k_{3}=0$  (Fig.~\ref{fig:lamz_lamy}(b)).
The results in Figs.~\ref{fig:lamz_lamy}(a) and (b) correspond to  Figs.~\ref{energylambda}(d) and (f) respectively.
Therefore, the results of this model calculation are consistent with the discussion in Sec.~\ref{cuttingprocedure}.

We discuss here the results for $\lambda_{2}=0$ with various values of $k_{3}$.
States are interchanged between the conduction and the valence bands in Fig.~\ref{fig:lamz_lamy}(c) ($k_{3}=0$), 
but not in Fig.~\ref{fig:lamz_lamy}(d) ($k_{3}=0.05$) and (e) ($k_{3}=0.1$). 
The result in Fig.~\ref{fig:lamz_lamy}(c) corresponds to Fig.~\ref{fig:appearing_hinge}(b) and the results in
Figs.~\ref{fig:lamz_lamy}(d) and (e) correspond to Fig.~\ref{fig:appearing_hinge}(c).
Fig.~\ref{fig:lamz_lamy}(f) is the energy spectrum when $k_{3}=\pi$.
As mentioned in Sec.~\ref{subsec:Appearance of hinge states}, an  
interchange of states between the conduction and the valence bands does not occur.
From the above, we confirmed that all the results from the model calculations are consistent with the discussion in Sec.~\ref{topological index and cutting procedure}. 

%newsection
\section{Positions of hinge states}\label{sec:positionofhingestates}
In this section, we consider a crystal with a parallelepiped shape with its edges along $x_{i}$ axis (see Fig.~\ref{fig:location_hinge}) and discuss which hinges support gapless hinge states. 
From the previous discussion, we find that pairs of hinge states always appear in insulators with $\mathbb{Z}_{4}=2$ when $\lambda_{1}=\lambda_{2}=0$. 
Each pair consists of a hinge state with positive velocity and one with negative velocity which  are related by inversion symmetry as shown in Fig.~\ref{tuikahingefig}.
In addition, the number of the pairs is odd.
When the system size is very large, every state should be at one hinge among the four hinges facing each other (see Fig.~\ref{tuikahingefig}). From the inversion symmetry of the whole system, when one hinge state is at one hinge, the other hinge state should reside at the hinge which is facing diagonally with that hinge in Fig.~\ref{tuikahingefig}.
Therefore we conclude that gapless states appear at hinges of the system as shown in Fig.~\ref{fig:location_hinge}(a).
In general, in Fig.~\ref{fig:location_hinge}(a), an odd number of pairs of hinge states appear in two hinges facing each other (blue lines) and even number of pairs of hinge states appear at the other two hinges facing each other (red lines) because the total number of the pairs is odd.

We have found appearance of hinge states along $x_{3}$ direction by introducing the cutting procedure along $x_{1}$ and $x_{2}$ directions in the arguments so far.
We similarly find hinge states along the $x_{1}$ direction by introducing the cutting procedure along the $x_{2}$ and $x_{3}$ directions.
From this, we can consider two cases as shown in Figs.~\ref{fig:location_hinge}(b) and (c) as patterns of positions and directions where hinge states appear.
We furthermore consider hinge states in the $x_{2}$ direction by introduce cutting procedure along the $x_{1}$ and $x_{3}$ directions.
Then, we can consider four cases of Figs.~\ref{fig:location_hinge}(d-g).

%%fig
\begin{figure}
\includegraphics{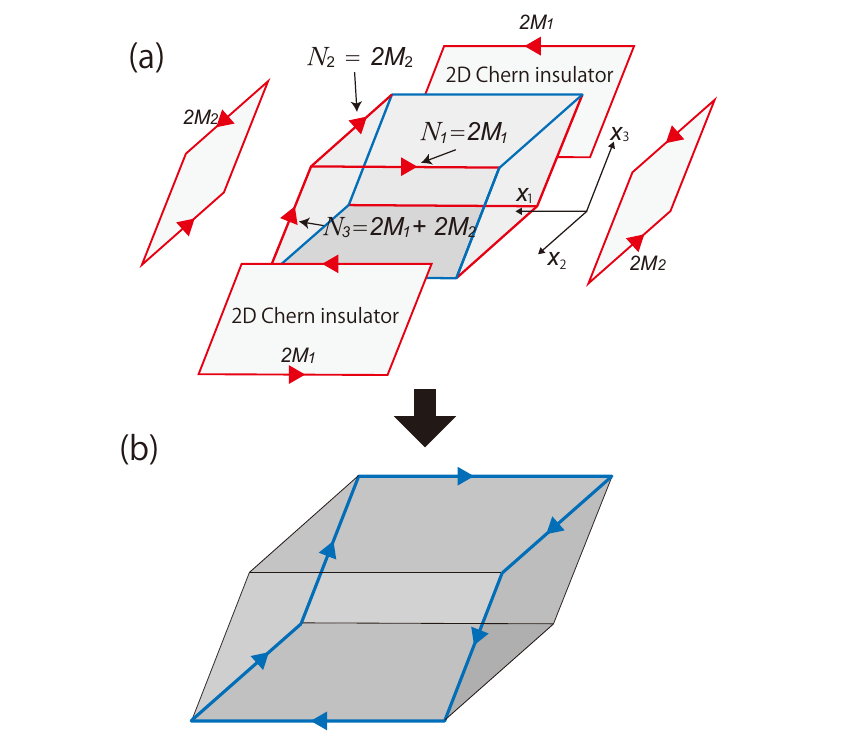}
\caption{\label{fig:hoti}Second-order topological insulator. (a) Let $N_{1}$, $N_{2}$ and $N_{3}$ be the number of $2M_{1}$, $2M_{2}$ and $2M_{1}+2M_{2}$ hinge states respectively.  
One can attach two 2D Chern insulators with same Chern number on two surfaces of opposite sides of the crystal, while preserving inversion symmetry. 
By attaching two 2D Chern insulators with Chern number $\mathcal{C}=2M_{1}$ and $\mathcal{C}=2M_{2}$ on $x_{1} x_{3}$-  and $x_{2}x_{3}$-surfaces respectively,
one can make the number of the hinge modes at the red hinges to be zero.
(b) Hinge states form a closed loop. In addition, the number of hinge states is odd.}
\end{figure}

Here we should discard unphysical cases among the four cases of Figs.~\ref{fig:location_hinge}(d-g).
In these figures, three hinges meet together at each corner  of the crystal. 
At each corner, the number of incoming hinge modes should be equal to that of outgoing hinge modes, where  ``incoming" and ``outgoing" refer to the signs of the velocities of hinge states.
It is shown as follows. In equilibrium, a current flows along the hinge modes, and at each corner the incoming current is equal to the outgoing current, because otherwise a charge will be accumulated at each corner in proportion with time. Then suppose we increase the chemical potential by $\Delta \mu$ within the gap. Each hinge mode will acquire an additional current by $\frac{e^{2}}{h}\Delta \mu$. For the current conservation at each corner after the shift $\Delta \mu$, the number of incoming hinge modes should be equal to that of outgoing hinge modes.
This argument is similar to the one in Ref.~\cite{berrypahaseinelectronic} for proving that the chiral edge currents are determined by a bulk orbital magnetization in 2D insulating ferromagnet.
From these discussions, we conclude that the only possible positions of hinge states are Figs.~\ref{fig:location_hinge}(d) and (g).
Because Fig.~\ref{fig:location_hinge}(g) is reduced to Fig.~\ref{fig:location_hinge}(d) by flipping the sign of $x_{3}$, Fig.~\ref{fig:location_hinge}(d) is essentially the only possibility for hinge modes.

In addition, we have some freedom in modifying the hinge modes without closing the bulk gap.
One can attach two-dimensional Chern insulators on the surfaces of the system, which modifies the number of hinge modes at each hinge while keeping the bulk unchanged. 
This discussion is similar to the one in Refs.~\cite{schindler2018higher,PhysRevB.97.205136,PhysRevB.98.205129}.
Note that this operation should preserve inversion symmetry. Therefore, we should simultaneously attach two 2D Chern insulators with the same Chern number on two surfaces of the opposite sides of the crystal. Then in Fig.~\ref{fig:hoti}, one can make the number of the hinge modes at the hinges with an even number of hinge modes (shown in red) to be zero.
To show this let us put $N_{1}=2M_{1}$ and $N_{2}=2M_{2}$ ($M_{1}, M_{2}$: integer), and we get $N_{3}=-2(M_{1}+M_{2})$. Then we attach 2D Chern insulators with Chern number $2M_{1}$ onto $x_{1} x_{3}$-surfaces, and those with Chern number $2M_{2}$ onto $x_{2}x_{3}$-surfaces. 
As a result, there is no longer a hinge state along the red lines, and remaing hinge gapless states form a closed loop as shown in Fig.~\ref{fig:hoti}(b).
In addition, the number of hinge modes is an odd number.

Thus, to summarize we have shown that the distribution of the gapleess hinge states is as shown in Fig.~\ref{fig:hoti}(b), by using the freedom to attach 2D Chern insulators while preserving inversion symmetry.
In previous papers \cite{,PhysRevB.97.205136,PhysRevB.98.205129}, the same distribution has been proposed for particular examples of SOTIs realized as $\mathbb{Z}_{2}$ topological insulators with magnetic field or magnetization. 
However, it is not obvious whether it holds for general SOTIs.
Here, we have shown that hinge states appear as shown in Fig.~\ref{fig:hoti}(b) when surfaces are gapped and $\mathbb{Z}_{4}=2$ without relying upon specific models.

\section{CONCLUSION}\label{section:conclusion}
In this paper, we give a general proof that any insulators with inversion symmetry and gapped surface always have hinge states when $\mathbb{Z}_{4}$ topological index $\mu_{1}$ is $\mu_{1}=2$.
In the proof, we introduce the cutting procedure. We change boundary conditions along two directions by changing hopping amplitudes across the boundaries, and study behaviors of gapless states through this change. 
We then reveal that the behaviors of gapless states result from the strong $\mathbb{Z}_{4}$ topological index. 
 From this discussion, we show that when the strong 
$\mathbb{Z}_{4}$ topological index $\mu_{1}$ is $\mu_{1}=2$ and the weak topological indices $\nu_{1}$, $\nu_{2}$ and $\nu_{3}$ are $\nu_{1}=\nu_{2}=\nu_{3}=0$, gapless states appear inevitably at the hinges of three-dimensional insulators with gapped surfaces.
We also identify the only possible configuration for the hinge modes as in Fig.~\ref{fig:location_hinge}(d).
Together with a freedom to attach 2D Chern insulators on surfaces, it can always be reduced to Fig.~\ref{fig:hoti}(b) with an odd number of chiral hinge states.

\begin{acknowledgments}
This work was supported by JSPS KAKENHI Grant numbers JP18H03678 and JP16J07354; by JST - CREST Grant number JP-MJCR14F1; and by the MEXT Elements Strategy Initiative to Form Core Research Center (TIES).  R. T. was also supported by JSPS KAKENHI Grant number JP18J23289.
\end{acknowledgments}

\appendix\
%%appendix A
\section{Proof of $E(-\lambda)=E(\lambda)$ for localized states with opposite parities\label{apendixa}}
In this appendix, we prove the following three propeties regarding the cutting procedure with the cutting parameter $\lambda$: (i) For the boundary localized states, the energy spectrum is symmetric with respect to the sign change of $\lambda$, i.e. $E(\lambda)=E(-\lambda)$. (ii) The boundary localized states $\ket{\psi (\lambda)}$ and $\ket{\psi (-\lambda)}$ have opposite-parity eigenvalues. (iii) Through a unitary transformation $U_{x}=\exp[i\pi \hat{x}/L]$, the Bloch wave vector is shifted as $k_{x}\rightarrow k_{x}+\pi/L$, and anti-periodic boundary condition is converted into periodic boundary condition. 

Here, we consider a one-dimensional periodic system with the coordinate $x$. Let the system size in $x$-direction be $L=2M+1$ with an integer $M$ measured in the unit of $|\boldsymbol{a}|$. The positions of the unit cells are represented by $x=-M,-M+1,\cdots M$.
For simplicity, at first, we only consider the case where the hopping is limited up to the nearest neighbor unit cells. Then the Hamiltonian is expressed as follows:
\begin{align}
\mathcal{H}(\lambda)&=
\begin{pmatrix}
H_0 & H_1 &  & \lambda H_1^{\dagger}\\
H_1^{\dagger} & H_0 & \ddots  &  \\
&  \ddots & \ddots & H_1 \\
\lambda H_1&  & H_1^{\dagger} & H_0
\end{pmatrix}\nonumber \\
&=\sum_{x=-M}^{M} H_{0} \otimes \ket{x}\bra{x}\nonumber \\
&\ \ \ \ \ \ +\sum_{x=-M}^{M-1}\bigl(H_{1}\otimes \ket{x+1}\bra{x}+\rm{H.c.}\bigr) \nonumber \\
&\ \ \ \ \ \ + \bigl( \lambda H_{1}\otimes \ket{-M}\bra{M} +\rm{H.c.} \bigr).
\end{align}
Here, $H_{0}$ and $H_{1}$ are $N_{0}\times N_{0}$ matrices, where $N_{0}$ is the number of states at each unit cell, coming from internal degrees of freedom. $H_{0}$ and $H_{1}$ represent the intra unit-cell term and the nearest neighbor hopping term, respectively.
Note that this model can be easily extended to two- and three-dimensional systems by adding the matrices representing hoppings along the $y$ and $z$ directions to $H_{0}$ and $H_{1}$ as internal degrees of freedom. 

\subsection{Proof of $E(-\lambda)=E(\lambda)$ for localized states}\label{appendixa1proofofeforlofalized} 
Here, we define the projection operator $P^{(l)}$ as follows: 
\begin{align}
P^{(l)}=
\begin{pmatrix}
\mathbf{1} & & & & & & & & \\
 & \ddots  & & & & & & & \\
 & & \mathbf{1} & \mathbf{0} & \cdots & \mathbf{0} & & & \\
 & & & \mathbf{0} & \cdots & \mathbf{0} & \mathbf{1} & & \\
 & & & & & & & \ddots & \\
 & & & & & & & & \mathbf{1} 
\end{pmatrix},
\end{align}
where $\mathbf{1}$ is a $N_{0}\times N_{0}$ identity matrix, and $\mathbf{0}$ is a $N_{0}\times N_{0}$ zero matrix.
Here, $P^{(l)}$ is a $2lN_0\times LN_0$ matrix, where 
%$L$ is a number of unit cell, $N_0$ is a number of internal degrees of freedom, and 
$l$ is taken as the penetration depth of the localized states. 
We define the projected Hamiltonian as follows: 
\begin{align}
\mathcal{H}^{(l)}(\lambda)
=P^{(l)}\mathcal{H}(\lambda)(P^{(l)})^{\dagger}. 
\end{align}
Now, we assume that the boundary localized state is well described by the projected Hamiltonian. In order to show that the energy spectrum of the localized state is symmetric, it is sufficient to show that that $\mathcal{H}^{(l)}(\lambda)$ and $\mathcal{H}^{(l)}(-\lambda)$ have same energy spectrum. In the following, we show this for the general $l$. 

As a simple example, we first consider the $l=1$ case. 
When $l=1$, the projected Hamiltonian is calculated as follows: 
\begin{align}
&\mathcal{H}^{(1)}(\lambda)
=P^{(1)}\mathcal{H}(\lambda)(P^{(1)})^{\dagger}
\notag \\
=&
\begin{pmatrix}
\mathbf{1} & \mathbf{0} & \cdots & \mathbf{0} \\
\mathbf{0} & \cdots & \mathbf{0} & \mathbf{1} 
\end{pmatrix}
%\mathcal{H}(\lambda)
\begin{pmatrix}
H_0 & H_1 &  & \lambda H_1^{\dagger}\\
H_1^{\dagger} & H_0 & \ddots  &  \\
&  \ddots & \ddots & H_1 \\
\lambda H_1&  & H_1^{\dagger} & H_0
\end{pmatrix}
\begin{pmatrix}
\mathbf{1} & \mathbf{0} \\
\mathbf{0} & \vdots \\
\vdots & \mathbf{0} \\
\mathbf{0} & \mathbf{1} \\
\end{pmatrix}
\notag \\
=&
\begin{pmatrix}
H_0 &  \lambda H_1^{\dagger}\\
\lambda H_1& H_0
\end{pmatrix}.
\end{align}
We can show that $\mathcal{H}^{(1)}(\lambda)$ and $\mathcal{H}^{(1)}(-\lambda)$ have same energy spectrum. We can easily check the following relation holds:
\begin{align}
\mathcal{H}^{(1)}(-\lambda)
=
\begin{pmatrix}
\mathbf{1} & \mathbf{0}\\
\mathbf{0} & -\mathbf{1}
\end{pmatrix}
\mathcal{H}^{(1)}(\lambda)
\begin{pmatrix}
\mathbf{1} & \mathbf{0}\\
\mathbf{0} & -\mathbf{1}
\end{pmatrix}. \label{Hm_SHS}
\end{align}
%Therefore, by taking the determinant of both sides of Eq. (\ref{Hm_SHS}), we conclude that $\text{det}(\mathcal{H}^{(1)}(\lambda))=\text{det}(\mathcal{H}^{(1)}(-\lambda))$,
Therefore, $\mathcal{H}^{(1)}(\lambda)$ and $\mathcal{H}^{(1)}(-\lambda)$ are unitary equivalent, and have same energy spectrum. 

We then show that similar discussion holds true for general $l$. 
In general, $\mathcal{H}^{(l)}(\lambda)$ have the following form: 
\begin{align}
\mathcal{H}^{(l)}(\lambda)
=
\begin{pmatrix}
X &  \lambda Y^{\dagger}\\
\lambda Y & X
\end{pmatrix}.
\end{align}
Here, $X$ and $Y$ are $lN_0 \times lN_0$ matrix shown as follows:
\begin{align}
X&=
\underbrace{
\begin{pmatrix}
H_0 & H_1 &  & \\
H_1^{\dagger} & H_0 & \ddots  &  \\
&  \ddots & \ddots & H_1 \\
& & H_1^{\dagger} & H_0
\end{pmatrix},
}_{l\ \text{blocks}}
\\
Y&=
\underbrace{
\begin{pmatrix}
\mathbf{0} & &  & \\
\vdots & \ddots & &  \\
\mathbf{0} & & \ddots & \\
H_1 & \mathbf{0} & \cdots & \mathbf{0}
\end{pmatrix}.
}_{l\ \text{blocks}}
\end{align}
$\mathcal{H}^{(l)}(\lambda)$ and $\mathcal{H}^{(l)}(-\lambda)$ are unitary equivalent,
\begin{align}
\mathcal{H}^{(l)}(-\lambda)
=
U_{l}
\mathcal{H}^{(l)}(\lambda)
U_{l}^{\dagger},
\end{align}
where $U_{l}$ is an unitary operator, defined as
\begin{align}
U_{l}=
\begin{pmatrix}
\mathbf{1}_{lN_0} & \\
 & -\mathbf{1}_{lN_0}
\end{pmatrix}.
\end{align}
and $\mathbf{1}_{lN_0}$ is an $lN_0 \times lN_0$ identity matrix.
If $\ket{\psi (\lambda)}$ is an eigenstate of $\mathcal{H}^{(l)} (\lambda)$, $U_{l}\ket{\psi (\lambda)}$ is an eigenstate of $\mathcal{H}^{(l)}(-\lambda)$. That is because
\begin{align}
\mathcal{H}^{(l)}(-\lambda)U_{l}\ket{\psi (\lambda)}&=U_{l}
\mathcal{H}^{(l)}(\lambda) U_{l}^{\dagger} U_{l} \ket{\psi (\lambda)}\nonumber \\
&=U_{l} \mathcal{H}^{(l)}(\lambda) \ket{\psi (\lambda)} \nonumber \\
&=E(\lambda) U_{l} \ket{\psi (\lambda)}.
\end{align}
Therefore, $\mathcal{H}^{(l)}(\lambda)$ and $\mathcal{H}^{(l)}(-\lambda)$ have the same energy spectrum. 
In reality, the localized states at the boundary have exponential tails into the bulk, and they are not strictly restricted within a finite number of sites in a thermodynamic limit.
Nonetheless, by extending the discussion on $\mathcal{H}^{(l)}(\lambda)$ to a larger value of $l$, one can see that $\mathcal{H}(\lambda)$ and $\mathcal{H}(-\lambda)$ have asymptotically the same spectra for localized states, when the system size becomes large.
We note that for delocalized states this proof of $E(\lambda)=E(-\lambda)$ is not valid, as we mentioned at the end of Sec.~\ref{subsec:Appearance of hinge states}.

\subsection{Localized states $\ket{\psi (\lambda)}$ and $\ket{\psi (-\lambda)}$ with opposite parity}
Here, we show that the boundary localized states $\ket{\psi(\lambda)}$ and $\ket{\psi(-\lambda)}$ have opposite parity eigenvalues. 
$\mathcal{H}^{(l)}(\lambda)$ is written in the basis consisting of $2lN_{0}$ states, located at $2l$ unit cells at $x=M, M-1, \cdots M-l+1, -M+l-1, \cdots -M+1, -M$ with $N_{0}$ representing internal degrees of freedoms. 
We define the inversion operator $I$ for the eigenstate $\ket{\psi (\lambda)}$ of $\mathcal{H}^{(l)}(\lambda)$, where the inversion center for $I$ is $x=0$.
By the unitary operator $U_{l}$, states at unit cells from $x=M$ to $x=M-l+1$ are multiplied by $1$ and those at unit cells from $x=-M+l-1$ to $x=-M$ are multiplied by $-1$, if the system size $L$ is sufficiently large. Therefore, we obtain the following relation.
\begin{equation}
U_{l}IU_{l}^{\dagger}=-I.
\end{equation}
From this relation, if $\ket{\psi (\lambda)}$ have parity eigenvalue $\xi$ of the inversion operator $I$, we can show that  $U_{l}\ket{\psi (\lambda)}$ have parity eigenvalue $-\xi$ because
\begin{align}
I U_{l} \ket{\psi (\lambda)}&=-U_{l}IU_{l}^{\dagger}U_{l}\ket{\psi (\lambda)}\nonumber \\
&=-U_{l}I\ket{\psi (\lambda)}\nonumber \\
&=-\xi U_{l} \ket{\psi (\lambda)}.
\end{align}
Because $U_{l}\ket{\psi (\lambda)}$ is the eigenstate of $\mathcal{H}(-\lambda)$, we conclude that $\ket{\psi(\lambda)}$ and $\ket{\psi(-\lambda)}$ have opposite parity eigenvalues.

\subsection{Unitary operator $U_{x}$ and anti-periodic boundary condition}
Here, we show that wave vector is shifted as $k_{x}\rightarrow k_{x}+\pi/L$ through the unitary transformation $U_{x}=\exp[i\pi \hat{x}/L]$, and thereby the anti-periodic boundary condition ($\lambda=-1$) is converted into the periodic boundary condition ($\lambda=1$). 
This unitary transformation $U_{x}$ is asymptotically equal to $U_{l}$ if the system size is sufficiently large and states are localized at the boundaries. 
The translational operator $T_{x}$ is defined as $T_{x}=\sum_{x=-M}^{M-1}\ket{x+1}\bra{x}+\ket{-M}\bra{M}$
. For $\lambda=1$, the Hamiltonian is expressed as follows:
\begin{align}\label{mathcalHlambda1}
&\ \ \ \ \ \mathcal{H}(\lambda=1)\nonumber \\
&=\sum_{x=-M}^{M} H_{0} \otimes \ket{x}\bra{x}+\sum_{x=-M}^{M-1}\bigl(H_{1}\otimes \ket{x+1}\bra{x}+\rm{H.c.}\bigr) \nonumber \\
&\ \ \ \ \ \ + \bigl( H_{1}\otimes \ket{-M}\bra{M} +\rm{H.c.} \bigr) \nonumber \\
&=H_{0}\otimes \Bigl( \sum_{x=-M}^{M}\ket{x}\bra{x}\Bigr)+\bigr[ H_{1}\otimes T_{x}+\rm{H.c.} \bigr].
\end{align}
For $\lambda=-1$, $U_{x}\mathcal{H}(\lambda=-1)U^{\dagger}_{x}$ is expressed as follows:
\begin{align}\label{UxmathcalHlambda-1Udaggerx}
&\ \ U_{x}\mathcal{H}(\lambda=-1)U^{\dagger}_{x}\nonumber \\
=&\sum_{x=-M}^{M} H_{0}\otimes \ket{x}\bra{x}e^{i\frac{\pi}{L}(x-x)}\nonumber \\
&+\sum_{x=-M}^{M-1}\Bigl( H_{1}\otimes \ket{x+1}\bra{x}e^{i\frac{\pi}{L}(x+1-x)}+\rm{H.c.}\Bigr)\nonumber \\
&-\Bigl( H_{1}\otimes \ket{-M}\bra{M}e^{i\frac{\pi}{L}((-M)-M)}+ \rm{H.c.} \Bigr)\nonumber \\
=&H_{0}\otimes \sum_{x=-M}^{M}\ket{x}\bra{x}\nonumber \\
&+\biggl[\bigl(H_{1}e^{i\frac{\pi}{L}}\bigr)\otimes \biggl(\sum_{x=-M}^{M-1}\ket{x+1}\bra{x} + \ket{-M}\bra{M}  \biggr) +\rm{H.c.}\biggr]\nonumber \\
=&H_{0}\otimes \Bigl( \sum_{x=-M}^{M}\ket{x}\bra{x}\Bigr)+\bigr[ (H_{1}e^{i\frac{\pi}{L}})\otimes T_{x}+\rm{H.c.} \bigr]\nonumber \\
=&H_{0}\otimes \Bigl( \sum_{x=-M}^{M}\ket{x}\bra{x}\Bigr)+\bigr[ H_{1}\otimes \tilde{T}_{x}+\rm{H.c.} \bigr],
\end{align}
where $\tilde{T}_{x}=e^{i\frac{\pi}{L}}T_{x}$. Let $e^{ik}$, $e^{i\tilde{k}}$ be eigenvalues of $T_{x}$, $\tilde{T}_{x}$ respectively. $k$, $\tilde{k}$ take values as follows:
\begin{equation}
k=\frac{2\pi}{L}m\ \ \ (-M\leq m \leq M),
\end{equation}
\begin{equation}
\tilde{k}=\frac{2\pi}{L}m+\frac{\pi}{L}\ \ \ (-M\leq m \leq M).
\end{equation}
By comparing Eqs.~(\ref{mathcalHlambda1}) and (\ref{UxmathcalHlambda-1Udaggerx}), we conclude that $U_{x}\mathcal{H}(\lambda=-1) U^{\dagger}_{x}$ is unitary equivalent to $\mathcal{H}(\lambda=1)$ with $k$ shifted to $k+\pi/L$. Therefore we conclude that,  through the unitary transformation $U_{x}$,  the Bloch wave vector is shifted as $k\rightarrow k+\frac{\pi}{L}$ and the anti-periodic boundary condition is converted into the periodic boundary condition.

%%appendixB
\section{\label{section:extension to the general cases}Proof of the existence of hinge states, and general combinations of parity eigenvalues at TRIM}
In the main text, we have considered the case when there are two odd-parity eigenstates at $\Gamma$ point and no odd-parity states at the other TRIM. In addition we have assumed that the number of occupied bands is two in the main text.
In this appendix, we extend our theory to an arbitrary number of occupied bands and general combinations of parity eigenvalues at TRIM, while keeping the strong $\mathbb{Z}_{4}$ index $\mu_{1}=2$ and weak $\mathbb{Z}_{2}$ indices $\nu_{1}=\nu_{2}=\nu_{3}=0$.
\subsection{Parity eigenvalues at TRIM in general cases}\label{seubsec:paritygeneralcases}
%fig
\begin{figure}
\includegraphics{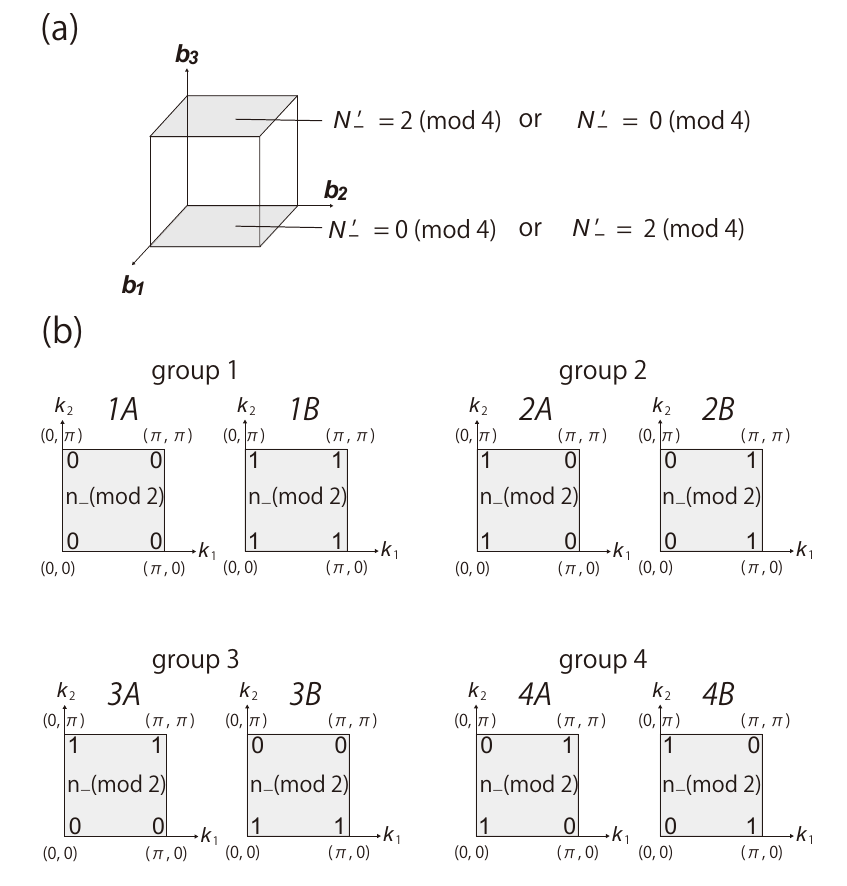}
\caption{\label{fig:general}General cases of parity eigenvalues at TRIM. 
(a) We divide eight TRIM $\Gamma_{j}=(n_{1}\boldsymbol{b}_{1}+n_{2}\boldsymbol{b}_{2}+n_{3}\boldsymbol{b}_{3})/2$ into two planes with $n_{3}=0$ and $n_{3}=1$. We can show that $N^{\prime}_{-}=2$ for one plane and $N^{\prime}_{-}=0$ for the other plane. (b) The number $n_{-}(k_{1},k_{2},n_{3})$ modulo 2 at TRIM on the planes $n_{3}=0$ and  $n_{3}=1$. Only combinations within the same group are allowable for the two planes $n_{3}=0$ and $n_{3}=1$.
For example, if 1A is realized on the plane $n_{3}=0$, only 1A and 1B are allowed on the plane $n_{3}=1$.}
\end{figure}
%fig
Here, we consider general combinations of parity eigenvalues at TRIM.
First, we note that the total number of odd-parity eigenstates at TRIM is 2 (mod 4) because the strong $\mathbb{Z}_{4}$ index $\mu_{1}=2$ (mod 4).
As shown in Fig.~\ref{fig:general}(a), 
we divide the eight TRIM $\Gamma_{j}=(n_{1}\boldsymbol{b}_{1}+n_{2}\boldsymbol{b}_{2}+n_{3}\boldsymbol{b}_{3})/2$ into two planes $n_{3}=0$ and $n_{3}=1$.
Here, let $N^{\prime}_{-}(n_{3})$ be the total numbers of odd-parity eigenstates at four TRIM on the plane $n_{3}={\rm const}\ (=0,1)$.

We can consider many cases of combinations of parity eigenvalues when $\mu_{1}=2$.
From the strong index $\mu_{1}=2$,
the possible combinations of $(N^{\prime}_{-}(n_{3}=0), N^{\prime}_{-}(n_{3}=1))$ are $(0,2)$, $(2,0)$, $(1,1)$ and $(3,3)$  modulo 4.
However, the weak index is $\nu_{3}=1$ in the cases $(1,1)$ and $(3,3)$, and these cases should be excluded.
In this way, the combinations of parity eigenvalues are restricted by the condition $\nu_{3}=0$. Thus we obtain the following relation.
\begin{align}\label{nminushazerotwotwozero}
\bigl( N^{\prime}_{-}(n_{3}=0),N^{\prime}_{-}(n_{3}=1)\bigr)
=\begin{cases} \bigl(0,2\bigr) \ ({\rm mod}\ 4)\\
\bigl(2,0\bigr)\ ({\rm mod}\ 4),
\end{cases}
\end{align}
as shown in  Fig.~\ref{fig:general}(a).

Next, let $n_{-}(k_{1},k_{2},n_{3})$ be the number of odd-parity eigenstates at four TRIM for a fixed value of $n_{3}=0$ or $n_{3}=1$, where $k_{1}$ and $k_{2}$ are wave-vectors along the $\boldsymbol{b_{1}}$ and $\boldsymbol{b_{2}}$ directions. 
There are eight patterns as combinations of four $n_{-}(k_{1},k_{2},n_{3})$ (mod 2) at TRIM on plane of $n_{3}=0$ or $n_{3}=1$ as shown in Fig.~\ref{fig:general}(b) because $N^{\prime}_{-}(n_{3})=0$ and $2$ modulo 4.

Let us consider combinations of these patterns for the planes $n_{3}=0$ and $n_{3}=1$.
Since the weak indices are $\nu_{1}=0$ and $\nu_{2}=0$, the combinations of the patterns on the plane $n_{3}=0$ and the plane $n_{3}=1$ are restricted.
From this, the eight patterns can be classified into four groups as shown in Fig.~\ref{fig:general}(b), and
only combinations within the same groups are allowable for the combinations of the two planes $n_{3}=0$ and $n_{3}=1$.
For example, if the pattern 1A in group 1 is selected on one plane and the pattern 4A in group 4 on the other plane in Fig.~\ref{fig:general}(b),
the weak indices become $\nu_{1}=1$ and $\nu_{2}=1$, which  contradicts our assumption.
After all,
in order to satisfy $\nu_{1}=\nu_{2}=0$,
only combinations within the same groups are allowable.
Note that the model considered in section \ref{topological index and cutting procedure} and \ref{section:Tight-binding model} are classified as 1A both on $n_{3}=0$ and $n_{3}=1$ in Fig. \ref{fig:general}(b).

Here we define a quantity $\delta N(n_{3}) \equiv n_{-}(\pi, \pi,n_{3}) +n_{-}(0, \pi,n_{3})-n_{-}(\pi, 0, n_{3}) -n_{-}(0, 0,n_{3})$, which will be directly related to hinge states later.
$\delta N(n_{3})$ represent the difference between the number of odd parity at TRIM with $k_{3}=0$ and $k_{3}=\pi$. 
Because $N^{\prime}_{-}(n_{3})=n_{-}(\pi, \pi,n_{3}) +n_{-}(0, \pi,n_{3})+n_{-}(\pi, 0,n_{3}) +n_{-}(0, 0,n_{3})$, $\delta N(n_{3})$ can be expressed as follows:
\begin{align}\label{eq:dnkone}
\delta N(n_{3}) =2n_{-}(\pi, \pi,n_{3})+2n_{-}(0, \pi,n_{3})-N^{\prime}_{-}(n_{3}).
\end{align}
In addition, for all the groups in Fig. \ref{fig:general}(b), we have
\begin{align}\label{eq:group1234}
&\ 2n_{-}(\pi, \pi,n_{3})+2n_{-}(0, \pi, n_{3})\nonumber \\=&\begin{cases}
0 \ \ ({\rm mod}\ 4) & {\rm for\ group}\ 1\\
2 \ \ ({\rm mod}\ 4) & {\rm for\ group}\ 2\\
0 \ \ ({\rm mod}\ 4) & {\rm for\ group}\ 3\\
2 \ \ ({\rm mod}\ 4) & {\rm for\ group}\ 4.
\end{cases}
\end{align}
Therefore $\delta N(n_{3})$ can be expressed as follows:
\begin{align}\label{deltan}
\delta N(n_{3})=\begin{cases}
N^{\prime}_{-}(n_{3}) \ \ \ \ \ \ \ ({\rm mod}\ 4) & {\rm for\ group}\ 1\\
N^{\prime}_{-}(n_{3})-2 \ \ ({\rm mod}\ 4) & {\rm for\ group}\ 2\\
N^{\prime}_{-}(n_{3}) \ \ \ \ \ \ \ ({\rm mod}\ 4) & {\rm for\ group}\ 3\\
N^{\prime}_{-}(n_{3})-2 \ \ ({\rm mod}\ 4) & {\rm for\ group}\ 4,
\end{cases}
\end{align}
where we need $N^{\prime}_{-}(n_{3})=0$ or 2 (mod 4).
As described below, this result is important. From Eqs.~(\ref{nminushazerotwotwozero}) and (\ref{deltan}), on the two places $n_{3}=0$ and $n_{3}=1$, $\delta N(n_{3})=2$ on  one plane and $\delta N(n_{3})=0$ on the other plane; this is related to the appearance of hinge states.
\subsection{Cutting procedure in the general cases of parity eigenvalues at TRIM}
%fig
\begin{figure}
\includegraphics{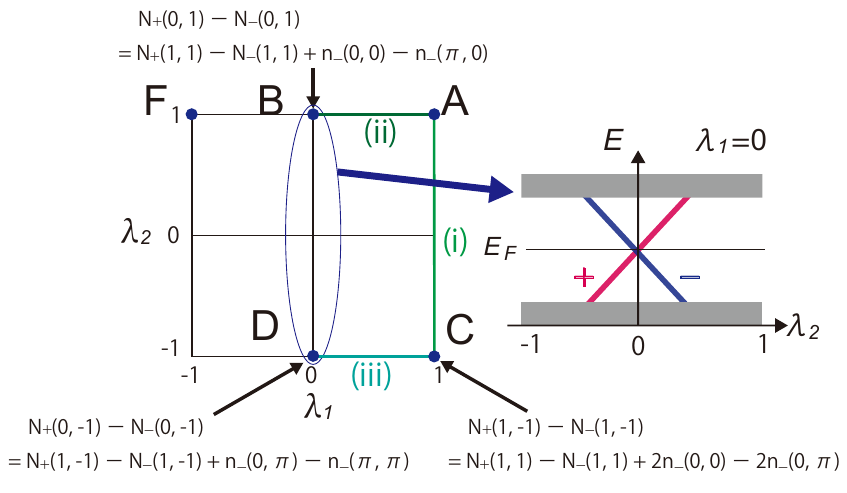}
\caption{\label{fig:hinge_general}(Color online) The states with even- and odd-parity eigenvalues are interchanged by varying $\lambda_{2}$ from $\lambda_{2}=1$ to $-1$ with $\lambda_{1}=0$ because of the difference between the values of $N_{+-}(\lambda_{1},\lambda_{2},n_{3})$ at the points B and D.}
\end{figure}
%fig
Here, we consider the cutting procedure again. 
Let $N_{+}(\lambda_{1},\lambda_{2},n_{3})$ and $N_{-}(\lambda_{1}, \lambda_{2},n_{3})$ be the numbers of  even- and odd-parity eigenstates below the Fermi level in a system with the cutting parameters $\lambda_{1}$ and $\lambda_{2}$.
In this subsection, we show that $\delta N(n_{3})$ is equal to the change in $N_{+-}(\lambda_{1},\lambda_{2},n_{3})\equiv N_{+}(\lambda_{1},\lambda_{2},n_{3})-N_{-}(\lambda_{1},\lambda_{2},n_{3})$ when $\lambda_{2}$ varies from the point B to the point D in Fig.~\ref{fig:hinge_general} (see Eq.~(\ref{eq:conclusionone})).
For this purpose, we calculate $N_{+-}(\lambda_{1},\lambda_{2},n_{3})$ at the point (i) C, (ii) B and (iii) D in Fig.~\ref{fig:hinge_general} in order.
Finally, we calculate the change in $N_{+-}(\lambda_{1},\lambda_{2},n_{3})$ when $\lambda_{2}$ varies from the point B to the point D. 

(i) First we calculate the change in $N_{+-}(\lambda_{1},\lambda_{2},n_{3})$ from the point A to the point C in Fig.~\ref{fig:hinge_general}.
Therefore, we consider the energy spectra when $\lambda_{1}=1$ and $k_{3}=0$ or $\pi$ are fixed and $\lambda_{2}$ varies.
Here we conclude that in changing from $\lambda_{2}=1$ to $\lambda_{2}=-1$, $N_{+-}(\lambda_{1}=1,\lambda_{2},n_{3})$ is increased by $2n_{-}(0,0,n_{3})-2n_{-}(0,\pi,n_{3})$:
\begin{align}\label{eq:difference}
&\bigl[ N_{+-}(\lambda_{1}=1,\lambda_{2}, n_{3}) \bigr]^{\lambda_{2}=-1}_{\lambda_{2}=1}\nonumber \\
=&2n_{-}(0,0,n_{3})-2n_{-}(0,\pi,n_{3}).
\end{align}
It is because in changing from $\lambda_{2}=1$ to $\lambda_{2}=-1$, $n_{-}(0,0,n_{3})-n_{-}(0,\pi,n_{3})$ states with odd parity shift from the valence bands to the conduction bands, and $n_{-}(0,0,n_{3})-n_{-}(0,\pi,n_{3})$ states with even parity shift from the conduction bands to the valence bands.

(ii) Next, we calculate the change in $N_{+-}(\lambda_{1},\lambda_{2},n_{3})$  from the point A to the point B in Fig.~\ref{fig:hinge_general}.
First, the change in $N_{-}(\lambda_{1},\lambda_{2},n_{3})$ from the point A to the point F is equal to $n_{-}(0,0,n_{3})-n_{-}(\pi,0,n_{3})$. Then,
we obtain the following relation.
\begin{align}\label{eq:differenceone}
&\bigl[ N_{+-}(\lambda_{1},\lambda_{2}=1,n_{3}) \bigr]^{\lambda_{1}=0}_{\lambda_{1}=1}\nonumber \\
=&\bigl[ N_{-}(\lambda_{1}, \lambda_{2}=1,n_{3})\bigr]^{\lambda_{1}=0}_{\lambda_{1}=-1}
-\bigl[N_{-}(\lambda_{1}, \lambda_{2}=1,n_{3})\bigr]^{\lambda_{1}=0}_{\lambda_{1}=1}\nonumber \\
=&\bigl[ N_{-}(\lambda_{1}, \lambda_{2}=1,n_{3})\bigr]^{\lambda_{1}=1}_{\lambda_{1}=-1}\nonumber \\
=&n_{-}(0,0,n_{3})-n_{-}(\pi,0,n_{3}),
\end{align}
where we use the following relation:
\begin{equation}
\bigl[ N_{\pm}(\lambda_{1}, \lambda_{2},n_{3}) \bigr]^{\lambda_{1}=0}_{\lambda_{1}=1}=
\bigl[ N_{\mp}(\lambda_{1}, \lambda_{2},n_{3}) \bigr]^{\lambda_{1}=0}_{\lambda_{1}=-1}.
\end{equation}
Eq.~(\ref{eq:differenceone}) represents the change in $N_{+-}(\lambda_{1},\lambda_{2},n_{3})$ from the point A to the point B.

(iii) From the same discussion, the change in $N_{+-}(\lambda_{1},\lambda_{2},n_{3})$ from the point C to the point D is expressed as follows:
\begin{align}\label{eq:differencetwo}
&\bigl[ N_{+-}(\lambda_{1}, \lambda_{2}=-1,n_{3}) \bigr]^{\lambda_{1}=0}_{\lambda_{1}=1}\nonumber \\
=&n_{-}(0,\pi,n_{3})-n_{-}(\pi,\pi,n_{3}).
\end{align}

From (i), (ii) and (iii), we are ready to calculate the change in $N_{+-}(\lambda_{1},\lambda_{2},n_{3})$ between the points B and D.
By combining Eqs.~(\ref{eq:difference}), (\ref{eq:differenceone}) and (\ref{eq:differencetwo}), we obtain the following result.
\begin{align}\label{ntasunhiku}
&\bigl[ N_{+-}(\lambda_{1}=0,\lambda_{2},n_{3})\bigr]^{\lambda_{2}=1}_{\lambda_{2}=-1}\nonumber \\
=&n_{-}(\pi,\pi,n_{3})+n_{-}(0,\pi,n_{3})\nonumber \\
&-n_{-}(\pi,0,n_{3})-n_{-}(0,0,n_{3}).
\end{align}
This is $\delta N(n_{3})$ defined in the previous subsection.
Therefore, from Eqs.~(\ref{deltan}) and (\ref{ntasunhiku}) we obtain the following relation:
\begin{align}\label{eq:conclusionone}
&\ \ \ \ \ \ \ \ \ \ \ \ \ \ \bigl[ N_{+-}(\lambda_{1}=0,\lambda_{2},n_{3}) \bigr]^{\lambda_{2}=1}_{\lambda_{2}=-1}\nonumber \\
=&\delta N(n_{3})=\begin{cases}
N^{\prime}_{-}(n_{3}) \ \ \ \ \ \ ({\rm mod}\ 4)\ {\rm group}\ 1\\
N^{\prime}_{-}(n_{3})-2 \ ({\rm mod}\ 4)\ {\rm group}\ 2\\
N^{\prime}_{-}(n_{3}) \ \ \ \ \ \ ({\rm mod}\ 4)\ {\rm group}\ 3\\
N^{\prime}_{-}(n_{3})-2 \ ({\rm mod}\ 4)\ {\rm group}\ 4.
\end{cases}
\end{align}
From Eq.~(\ref{nminushazerotwotwozero}), for any group in Eq.~(\ref{eq:conclusionone}), $\delta N(n_{3})=2$ for one of the two planes of $k_{3}=0$ and $k_{3}=\pi$, and $\delta N(n_{3})=0$ for the other plane.
This means that the eigenstates with even and odd parity are interchanged $1$ (mod 2) times for one of the two planes $k_{3}=0$ and $k_{3}=\pi$ by varying $\lambda_{2}$ from $\lambda_{2}=1$ to $-1$ when $\lambda_{1}=0$.
On the other plane, the interexchanges of eigenstates with even and odd parity occur $0$ (mod 2) times.
This is the same condition as in Sec.~\ref{subsec:Appearance of hinge states} of the main text, where we show that hinge states appear when the interexchange of states occurs when $k_{3}=0$ and the exchange does not occur when $k_{3}=\pi$.
Then, we conclude that hinge states appear for general cases when $\mu_{1}=2$ and $\nu_{1}=\nu_{2}=\nu_{3}=0$.
We also conclude that the number of the hinge states is odd, and the number of states across the Fermi level is odd from $k_{3}=0$ to $k_{3}=\pi$, from the same discussion as Sec.~\ref{subsec:Appearance of hinge states} of the main text.

%%appendixC
\section{Correspondence between symmetry-based indicators and hinge states in SOTIs with time-reversal symmetry}\label{section:timereversal}

\begin{figure}
\includegraphics{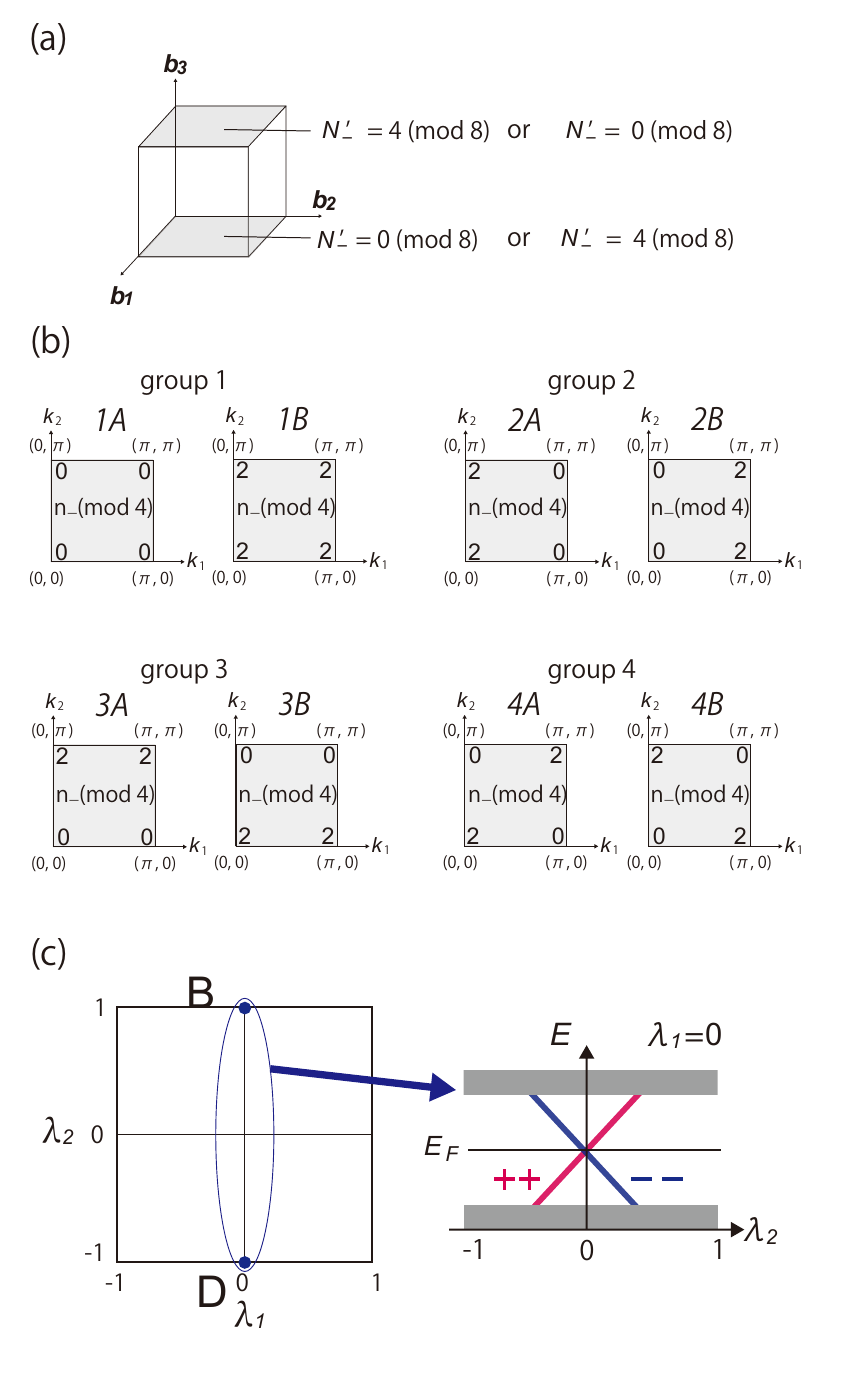}
\caption{\label{fig:timereversal}(Color online) Appearance of hinge states in SOTI with time-reversal symmetry.
(a) We divide eight TRIM $\Gamma_{j}=(n_{1}\boldsymbol{b}_{1}+n_{2}\boldsymbol{b}_{2}+n_{3}\boldsymbol{b}_{3})/2$ into two planes with $n_{3}=0$ and $n_{3}=1$. We can show that $N^{\prime}_{-}=4$ (mod 8) for one plane and $N^{\prime}_{-}=0$ (mod 8) for the other plane. (b) The number $n_{-}(k_{1},k_{2},n_{3})$ (mod 4) at TRIM on the planes $n_{3}=0$ and  $n_{3}=1$. Only combinations within the same group are allowable for each plane at constant $n_{3}$ ($=0,1$).
(c) Kramers pairs with even- and odd-parity eigenvalues are interchanged by varying $\lambda_{2}$ from the point B to the point D because of the difference between the values of $N_{+-}(\lambda_{1},\lambda_{2},n_{3})$ at the point B and the point D.}
\end{figure}
In the main text, we have considered a centrosymmetric system in class A. Here we consider a centrosymmetric system in class AII, that is, one with time-reversal symmetry.
For class AII, we can show that any insulators with inversion symmetry and with gapped surfaces always have hinge states when $\mathbb{Z}_{4}=2$ by introducing cutting procedure similarly to class A.
According to Refs.~\cite{po2017symmetry,PhysRevX.8.031070}, the symmetry-based indicator for class AII is found to be $X_{\rm BS}=\mathbb{Z}_{2}\times \mathbb{Z}_{2}\times \mathbb{Z}_{2}\times \mathbb{Z}_{4}$. Three $\mathbb{Z}_{2}$ factors are the weak topological indices, defined as
\begin{equation}
\nu^{\rm AII}_{a}\equiv \frac{1}{2}\sum_{\Gamma_{j}:{\rm TRIM} \land n_{a}=1} n_{-}(\Gamma_{j})\ \ ({\rm mod}\ 2),
\end{equation}
where $a=1,2,3$ and $n_{-}(\Gamma_{j})$ is the number of occupied states with odd parity at the TRIM $\Gamma_{j}$, and the summation is taken over the TRIM on the plane $n_{a}=1$..
In class AII systems, due to time-reversal symmetry, eigenstates at  the TRIM $\Gamma_{j}$ are Kramers-degenerate with the same parity eigenvalues.  Therefore, $n_{-}(\Gamma_{j})$ is an even number.
The $\mathbb{Z}_{4}$  factor is the strong topological index, defined as
\begin{align}
\kappa_{1}&=\frac{1}{4}\sum_{\Gamma_{j}: {\rm TRIM}}\Bigl(n_{+}(\Gamma_{j})-n_{-}(\Gamma_{j}) \Bigr)\ \  ({\rm mod}\ 4)\nonumber \\
&=-\frac{1}{2}\sum_{\Gamma_{j}:{\rm TRIM}}n_{-}(\Gamma_{j})\ \ ({\rm mod}\ 4),
\end{align}
where $n_{+}(\Gamma_{j})$ is the number of occupied states with even parity at the TRIM $\Gamma_{j}$. 
Therefore, in systems with inversion symmetry, topological phases are characterized by the symmetry-based indicator $X_{\rm BS}=(\nu_{1}^{\rm AII},\nu_{2}^{\rm AII},\nu_{3}^{\rm AII},\kappa_{1})$.
In the following, we show that hinge states appear when $(\nu_{1}^{\rm AII},\nu_{2}^{\rm AII},\nu_{3}^{\rm AII},\kappa_{1})=(0,0,0,2)$ using the discussion similar to that in Appendix~\ref{section:extension to the general cases}. 

Here, let $N^{\prime}_{-}(n_{3})$ be the total number of odd-parity eigenstates at four TRIM on a plane $n_{3}={\rm const}$ ($=0,1$).
First, we divide the eight TRIM $\Gamma_{j}$ into two planes $n_{3}=0$ and $n_{3}=1$  as shown in Fig.~\ref{fig:timereversal}(a) similarly to Appendix.~\ref{section:extension to the general cases}. 
For class AII, 
Eq.~(\ref{nminushazerotwotwozero}) is modified as follows:
\begin{align}\label{aiinminus04}
\bigl( N^{\prime}_{-}(n_{3}=0),N^{\prime}_{-}(n_{3}=1) \bigr)=\begin{cases} \bigl(0,4\bigr) \ ({\rm mod}\ 8)\\
\bigl(4,0\bigr)\ ({\rm mod}\ 8),
\end{cases}
\end{align}
because $\kappa_{1}=2$ and $\nu_{3}^{\rm AII}=0$. 

Next, let $n_{-}(k_{1},k_{2},n_{3})$ be the number of odd-parity eigenstates at four TRIM for a fixed value of $n_{3}=0$ or $n_{3}=1$. There are eight patterns as combinations of four $n_{-}(k_{1},k_{2},n_{3})$ at TRIM on plane of $n_{3}=0$ or $n_{3}=1$ as shown in Fig.~\ref{fig:timereversal}(b) because $N^{\prime}_{-}(n_{3})=0$ or 4 (mod 8).
In order to satisfy $\nu_{1}^{\rm AII}=\nu_{2}^{\rm AII}=0$, only combinations within the same groups are allowable as discussed in Appendix~\ref{section:extension to the general cases}.
From this,
Eq.~(\ref{eq:group1234}) is modified as follows:
\begin{align}\label{eq:group1234timereversal}
&\ 2n_{-}(\pi, \pi,n_{3})+2n_{-}(0, \pi, n_{3})\nonumber \\=&\begin{cases}
0 \ \ ({\rm mod}\ 8) & {\rm for\ group}\ 1\\
4 \ \ ({\rm mod}\ 8) & {\rm for\ group}\ 2\\
0 \ \ ({\rm mod}\ 8) & {\rm for\ group}\ 3\\
4 \ \ ({\rm mod}\ 8) & {\rm for\ group}\ 4.
\end{cases}
\end{align}
Therefore, Eq.~(\ref{deltan}) is also modified as follows:
\begin{align}\label{deltantimereversal}
\delta N(n_{3})=\begin{cases}
N^{\prime}_{-}(n_{3}) \ \ \ \ \ \ \ ({\rm mod}\ 8) & {\rm for\ group}\ 1\\
N^{\prime}_{-}(n_{3})-4 \ \ ({\rm mod}\ 8) & {\rm for\ group}\ 2\\
N^{\prime}_{-}(n_{3}) \ \ \ \ \ \ \ ({\rm mod}\ 8) & {\rm for\ group}\ 3\\
N^{\prime}_{-}(n_{3})-4 \ \ ({\rm mod}\ 8) & {\rm for\ group}\ 4,
\end{cases}
\end{align}
where $N^{\prime}_{-}(n_{3})=0$ or 4 (mod 8).
Eq.~(\ref{ntasunhiku}) holds true for class AII. Then, by combining Eqs.~(\ref{ntasunhiku}) and
(\ref{deltantimereversal}), we obtain the following relation:
\begin{align}\label{eq:conclusiononeaii}
&\ \ \ \ \ \ \ \ \ \ \ \ \ \ \ \bigl[ N_{+-}(\lambda_{1}=0,\lambda_{2},n_{3})\bigr]^{\lambda_{2}=1}_{\lambda_{2}=-1}\nonumber \\
=&\delta N(n_{3})=\begin{cases}
N^{\prime}_{-}(n_{3}) \ \ \ \ \ \ ({\rm mod}\ 8)\ {\rm group}\ 1\\
N^{\prime}_{-}(n_{3})-4 \ ({\rm mod}\ 8)\ {\rm group}\ 2\\
N^{\prime}_{-}(n_{3}) \ \ \ \ \ \ ({\rm mod}\ 8)\ {\rm group}\ 3\\
N^{\prime}_{-}(n_{3})-4 \ ({\rm mod}\ 8)\ {\rm group}\ 4.
\end{cases}
\end{align}
From Eq.~(\ref{aiinminus04}), for any group in Eq.~(\ref{eq:conclusiononeaii}), $\delta N(n_{{3}})=4$ for one of the two planes of $k_{3}=0$ and $k_{3}=\pi$, and $\delta N (n_{3})=0$ for the other plane. 
Therefore, Kramers pairs with even and odd parity are interchanged 1 (mod 2) times for one of the two planes $k_{3}=0$ and $k_{3}=\pi$ by varying $\lambda_{2}$ from $\lambda_{2}=1$ to $-1$ when $\lambda_{1}=0$ as shown in Fig.~\ref{fig:timereversal}(c).
On the other plane, the interexchanges of Kramers pairs with even and odd parity occur 0 (mod 2) times.
This discussion is the same as in Sec.~\ref{subsec:Appearance of hinge states}
when we replace a state in Sec.~\ref{subsec:Appearance of hinge states} with a Kramers pair.
Then, we conclude that hinge states appear when $\kappa_{1}=2$ and $\nu^{\rm AII}_{1}=\nu^{\rm AII}_{2}=\nu^{\rm AII}_{3}=0$ for class AII.
These hinge states are helical gapless states protected by time-reversal symmetry for class AII, while hinge states are chiral gapless states for class A.

%%appendixD
\section{Proof of existence of hinge states for a system with even numbers of the system sizes $L_{1}$ and $L_{2}$}\label{section:appendixd}
\begin{figure}
\includegraphics{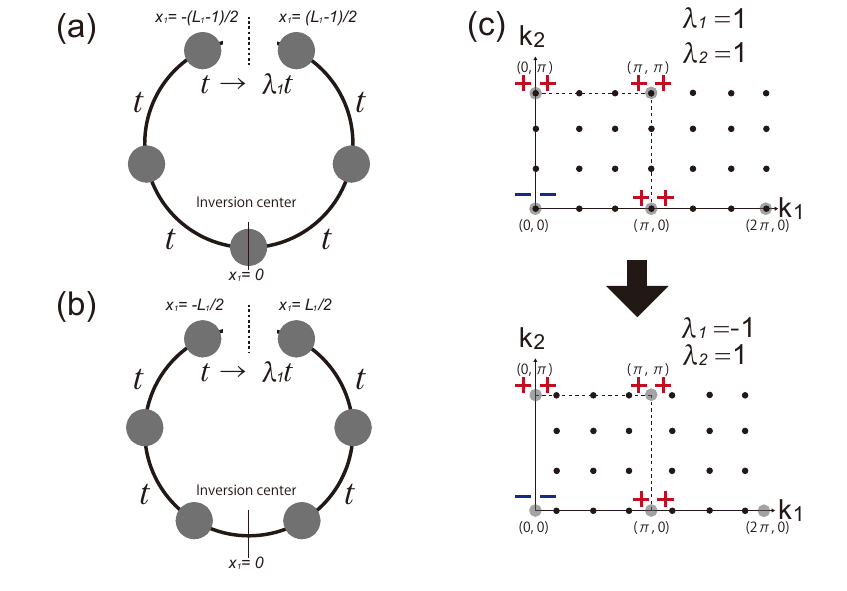}
\caption{\label{fig:appendixd} (Color online) One-dimensional systems with (a) odd values of the  system size $L_{1}$ and (b) even values of the system size $L_{1}$. (c) Black points represent possible wave vectors in the cases of periodic and anti-periodic boundary conditions in the $x_{1}$ direction when $L_{1}$ and $L_{2}$ are even. When $\lambda_{1}=1$ and $\lambda_{2}=1$, the four TRIM are among the possible wave vectors. On the other hand, no TRIM is among the possible wave vectors when $\lambda_{1}=-1$ and $\lambda_{2}=1$.}
\end{figure}
In the main text, we consider the system with odd numbers of the system size $L_{1}$ and $L_{2}$ in the $x_{1}$ and $x_{2}$ directions. In this appendix, we discuss the spectral flows via the cutting procedure in the case of a system with an even numbers of the system size $L_{i}$ ($i=1,2$). We set the system size in the $x_{3}$ direction as $L_{3}\rightarrow \infty$ and $k_{3}$ fixed to be $0$ or $\pi$. Other cases with $(L_{1},L_{2})=({\rm odd}, {\rm even})$ and $({\rm even}, {\rm odd})$ can be studied similarly, and we omit these cases here.

First, the inversion operators and the inversion centers are different between the case with odd and even numbers of $L_{i}$ ($i=1,2$). Therefore, it is necessary to distinguish the inversion operators in the even and odd cases. Then let $I_{\rm even}$ and $I_{\rm odd}$ be the inversion operators in the cases of even and odd values of $L_{i}$, respectively. For example, the inversion centers are different in one-dimensional systems with an odd and even values of $L_{1}$ as shown in Figs.~\ref{fig:appendixd}(a)(b). In the case of odd $L_{1}$, the inversion center $x=0$ is at the center of a unit cell. On the other hand, the inversion center is at the border between two neighboring unit cells in the case of even $L_{1}$. In one-dimensional systems, $I_{\rm even}$ is related to $I_{\rm odd}$ as follows.
\begin{equation}
I_{\rm even}=T_{x_{1}} I_{\rm odd},
\end{equation}
where $T_{x_{1}}$ is the translational operator in the $x_{1}$ direction. 
In two-dimensional systems with even $L_{1}$ and $L_{2}$, $I_{\rm even}$ is related to $I_{\rm odd}$ as follows similar to the one-dimensional case.
\begin{align}
&I_{\rm even}=T_{x_{1}}T_{x_{2}} I_{\rm odd},\\
&\xi_{\rm even}(k_{1}, k_{2})=e^{-ik_{1}}e^{-ik_{2}} \xi_{\rm odd}(k_{1},k_{2}),
\end{align}
where $\xi_{\rm even}(k_{1}, k_{2})$ and $\xi_{\rm odd}(k_{1}, k_{2})$ represent eigenvalues of $I_{\rm even}$ and $I_{\rm odd}$ respectively at $(k_{1},k_{2}) \in {\rm TRIM}$. From this, we find the following relations of $n_{\pm}(k_{1},k_{2},n_{3})$ in the case with even and odd $L_{i}$ $(i=1,2)$:
\begin{align}
&n^{\rm even}_{\pm}(0,0,n_{3})=n^{\rm odd}_{\pm}(0,0,n_{3}), \label{eventooddnorelation1}\\
&n^{\rm even}_{\pm}(\pi,0,n_{3})=\nu -n^{\rm odd}_{\pm}(\pi,0,n_{3}),\\
&n^{\rm even}_{\pm}(0,\pi, n_{3})=\nu - n^{\rm odd}_{\pm}(0,\pi , n_{3}),\\
&n^{\rm even}_{\pm}(\pi,\pi, n_{3})=n^{\rm odd}_{\pm}(\pi,\pi ,n_{3}),\label{eventooddnorelation4} 
\end{align}
where $\nu = n^{\rm odd}_{+}(k_{1},k_{2},n_{3})+n^{\rm odd}_{-}(k_{1},k_{2},n_{3})$.
When $(\lambda_{1},\lambda_{2})=(1,1)$, the Bloch wave-vector $k_{i}$ ($i=1,2$)  in the $x_{i}$ direction is
\begin{equation}
k_{i}=\frac{2\pi}{L_{i}} m_{i}\ \ \ \left(-\frac{L_{i}}{2}< m_{i}\leq \frac{L_{i}}{2}\right).
\end{equation}
Therefore, $(k_{1},k_{2})$ can take the values of the four TRIM $(0,0)$, $(\pi,0)$, $(0,\pi)$ and $(\pi, \pi)$ as shown in Fig.~\ref{fig:appendixd}(c). On the other hand, when $(\lambda_{1},\lambda_{2})=(-1,1)$, the set of 
the wave-vector $k_{2}$ does not change, and $k_{1}$ is given by 
\begin{equation}
k_{1}=\frac{2\pi}{L_{1}}m_{1}+\frac{\pi}{L_{1}}\ \ \ \left(-\frac{L_{i}}{2}< m_{i}\leq \frac{L_{i}}{2}  \right).
\end{equation}
From this, we find that $(k_{1},k_{2})$ cannot take TRIM. Each non-TRIM pair $(\boldsymbol{k},-\boldsymbol{k})$ with $k_{3}={\rm const}$ contributes 1 to  $N_{-}(\lambda_{1},\lambda_{2}, n_{3})$. From the above discussion, we obtain the following relations:
\begin{align}
N^{\rm even}_{-}(1,1, n_{3})=&\left( \frac{L_{1}L_{2}}{2}-2 \right)\nu \nonumber \\
&+n^{\rm even}_{-}(0,0,n_{3})
+n^{\rm even}_{-}(\pi,0,n_{3}) \nonumber \\
&+n^{\rm even}_{-}(0,\pi ,n_{3})+n^{\rm even}_{-}(\pi,\pi ,n_{3}), \nonumber \\
N^{\rm even}_{-}(-1,1, n_{3})=&\frac{L_{1}L_{2}}{2}\nu.
\end{align}
From these relations and Eqs.~(\ref{eventooddnorelation1}-\ref{eventooddnorelation4}), we obtain the following equation.
\begin{align}\label{evennotokinonnohenkasuuiti}
&[N^{\rm even}_{-}(\lambda_{1},\lambda_{2}=1,n_{3})]^{\lambda_{1}=1}_{\lambda_{1}=-1} \nonumber \\
=&n^{\rm odd}_{-}(0,0,n_{3})+n^{\rm odd}_{-}(\pi,\pi ,n_{3})\nonumber \\
&-\left(n^{\rm odd}_{-}(0,\pi ,n_{3})+n^{\rm odd}_{-}(\pi,0,n_{3})\right)
\end{align}
From similar discussion, the following equation holds.
\begin{align}
&[N^{\rm even}_{-}(\lambda_{1},\lambda_{2}=-1,n_{3})]^{\lambda_{1}=1}_{\lambda_{1}=-1}=0.
\end{align}

In Appendix~\ref{section:extension to the general cases}, we considered general combinations of parity eigenvalues at TRIM. In this case, the main result remains the same, but it is necessary to make some modifications. In the case of even $L_{i}$ ($i=1,2$), Eq.~(\ref{eq:difference}) in Appendix~\ref{section:extension to the general cases} is modified as follows:
\begin{align}\label{evencasebfive}
&[N^{\rm even}_{+-}(\lambda_{1}=1,\lambda_{2},n_{3})]^{\lambda_{2}=-1}_{\lambda_{2}=1}\nonumber \\
=&2n^{\rm odd}_{-}(0,0,n_{3})+2n^{\rm odd}_{-}(\pi,\pi,n_{3})\nonumber \\
&-\left(2n^{\rm odd}_{-}(0,\pi, n_{3})+2n^{\rm odd}_{-}(\pi,0,n_{3})\right).
\end{align}
In addition, Eq.~(\ref{eq:differenceone}) is modified as
\begin{align}\label{evencasebsix}
&[N^{\rm even}_{+-}(\lambda_{1},\lambda_{2}=1,n_{3})]^{\lambda_{1}=0}_{\lambda_{1}=1}\nonumber \\
=&n^{\rm odd}_{-}(0,0,n_{3})+n^{\rm odd}_{-}(\pi,\pi ,n_{3})\nonumber \\
&-\left(n^{\rm odd}_{-}(0,\pi ,n_{3})+n^{\rm odd}_{-}(\pi,0,n_{3})\right),
\end{align}
and Eq.~(\ref{eq:differencetwo}) is modified as
\begin{align}\label{evencasebeight}
&[N^{\rm even}_{+-}(\lambda_{1},\lambda_{2}=-1,n_{3})]^{\lambda_{1}=0}_{\lambda_{1}=1}
=0.
\end{align}
From Eqs.~(\ref{evencasebfive}-\ref{evencasebeight}), we obtain the following equation.
\begin{align}\label{mainresulteven}
&\bigl[ N^{\rm even}_{+-}(\lambda_{1}=0,\lambda_{2},n_{3})\bigr]^{\lambda_{2}=1}_{\lambda_{2}=-1}\nonumber \\
=&n^{\rm odd}_{-}(\pi,0,n_{3})+n^{\rm odd}_{-}(0,\pi,n_{3})\nonumber \\
&-n^{\rm odd}_{-}(0,0,n_{3})-n^{\rm odd}_{-}(\pi,\pi,n_{3})\nonumber \\
=&N'^{\rm odd}_{-}(n_{3})\nonumber \\
&-2\left(n^{\rm odd}_{-}(0,0,n_{3})+n^{\rm odd}_{-}(\pi,\pi,n_{3})\right),
\end{align}
where $N'^{\rm odd}_{-}(n_{3})$ is the total number of odd-parity eigenstates at four TRIM on a plane $n_{3}={\rm const}$ ($=0,1$) in the case with the odd system size.
Therefore, we obtain the following result for the four groups in Fig.~\ref{fig:general}:
\begin{align}
&\bigl[ N^{\rm even}_{+-}(\lambda_{1}=0,\lambda_{2},n_{3}) \bigr]^{\lambda_{2}=1}_{\lambda_{2}=-1}\nonumber \\
=&\begin{cases}
N^{\prime {\rm odd}}_{-}(n_{3}) \ \ \ \ \ \ ({\rm mod}\ 4)\ {\rm for\ group}\ 1\\
N^{\prime  {\rm odd}}_{-}(n_{3})-2 \ ({\rm mod}\ 4)\ {\rm for\ group}\ 2\\
N^{\prime  {\rm odd}}_{-}(n_{3})-2 \ ({\rm mod}\ 4)\ {\rm for\ group}\ 3\\
N^{\prime  {\rm odd}}_{-}(n_{3}) \ \ \ \ \ \ ({\rm mod}\ 4)\ {\rm for\ group}\ 4,
\end{cases}
\end{align}
where $N^{\prime  {\rm odd}}_{-}(n_{3})=0$ or 2 (mod 4).
From this, we find that for any group, the change of $N^{\rm even}_{+-}(\lambda_{1}=0,\lambda_{2},n_{3})$ in changing from $\lambda_{2}=1$ to $\lambda_{2}=-1$, is 2 (mod 4) for one of the two planes of $k_{3}=0$ and $k_{3}=\pi$. This change of $N^{\rm even}_{+-}(\lambda_{1}=0,\lambda_{2},n_{3})$ is  0 (mod 4) for the other plane. This is the same conclusion of Appendix~\ref{section:extension to the general cases} and the existence of gapless states is shown.
In Sec.~\ref{topological index and cutting procedure}, we consider the special case with $n^{\rm odd}_{-}(0,0)=2$ and $n^{\rm odd}_{-}(\pi,0)=n^{\rm odd}_{-}(0, \pi)=n^{\rm odd}_{-}(\pi,\pi)=0$. In this case, Eq.~(\ref{mainresulteven}) is equal to Eq.~(\ref{mainresult2b}).
Therefore, it is not necessary to modify the main result in Sec.~\ref{topological index and cutting procedure}, that is Eq.~(\ref{mainresult2b}) in the case with an even number of $L_{i}$ ($i=1,2$). 

\end{document}